\documentclass[aps,prx,reprint,superscriptaddress, amsmath, amssymb]{revtex4-2}
\usepackage{amsmath}
\usepackage{graphicx}
\usepackage{dcolumn}
\usepackage{bm}
\usepackage{braket}
\usepackage{ulem}
\usepackage[colorlinks]{hyperref}
\usepackage{subfigure}
\hypersetup{colorlinks=true, citecolor=blue, urlcolor=blue, linkcolor=blue}

\usepackage{mathtools}
\usepackage{xcolor}
\usepackage{xr}
\usepackage{graphicx}
\usepackage{lipsum} 
\usepackage{tabularx}
\usepackage{booktabs}%
\begin{document}

\title{Super-resolving frequency measurement with mode-selective quantum memory}

\author{Shicheng \surname{Zhang}}
\affiliation{Blackett Laboratory, Department of Physics, Imperial College London, Prince Consort Rd, London, SW7 2AZ, United Kingdom}
\author{Aonan \surname{Zhang}}
\email[]{aonan.zhang@physics.ox.ac.uk}
\affiliation{Blackett Laboratory, Department of Physics, Imperial College London, Prince Consort Rd, London, SW7 2AZ, United Kingdom}
\affiliation{Clarendon Laboratory, University of Oxford, Parks Road, Oxford OX1 3PU, United Kingdom}
\author{Ilse \surname{Maillette de Buy Wenniger}}
\affiliation{Blackett Laboratory, Department of Physics, Imperial College London, Prince Consort Rd, London, SW7 2AZ, United Kingdom}
\author{Paul M. \surname{Burdekin}}
\affiliation{Blackett Laboratory, Department of Physics, Imperial College London, Prince Consort Rd, London, SW7 2AZ, United Kingdom}
\author{Steven \surname{Sagona-Stophel}}
\affiliation{Blackett Laboratory, Department of Physics, Imperial College London, Prince Consort Rd, London, SW7 2AZ, United Kingdom}
\affiliation{Okinawa Institute of Science and Technology, Okinawa, Japan}
\author{Anindya \surname{Rastogi}}
\affiliation{Blackett Laboratory, Department of Physics, Imperial College London, Prince Consort Rd, London, SW7 2AZ, United Kingdom}
\author{Sarah E. \surname{Thomas}}
\affiliation{Blackett Laboratory, Department of Physics, Imperial College London, Prince Consort Rd, London, SW7 2AZ, United Kingdom}
\affiliation{Department of Engineering Science, University of Oxford, Parks Road, Oxford, OX1 3PJ, UK}
\author{Ian A. \surname{Walmsley}}
\affiliation{Blackett Laboratory, Department of Physics, Imperial College London, Prince Consort Rd, London, SW7 2AZ, United Kingdom}

\date{\today}

\begin{abstract}
High‑precision optical frequency measurement underpins modern science and technology, yet conventional spectroscopic techniques struggle to resolve sub-linewidth spectral features. Here, we introduce a platform for super‑resolved frequency estimation based on a mode‑selective atomic Raman quantum memory implemented in warm caesium vapour. By precisely engineering the light–matter interaction, the memory coherently stores the optimal temporal mode with high fidelity and retrieves it on demand, achieving mode crosstalk as low as 0.34\%. To estimate the separation between two spectral lines, we experimentally measure the mean squared error of the frequency estimate, reaching a sensitivity of 1/20 of the linewidth and a (34±4)-fold enhancement in precision over direct intensity measurements. This enhanced frequency resolution, combined with on‑demand storage, retrieval, and mode‑conversion capabilities, establishes a pathway toward multifunctional memory‑based time–frequency sensors and their integration within quantum networks.
\end{abstract}   

\maketitle

\section{Introduction} 
\par
The time-frequency (TF) degree of freedom of light underpins applications spanning high-resolution spectroscopy, precision timekeeping~\cite{Udem_2002,Ludlow_2015}, ultrafast optics~\cite{Walmsley:09}, and emerging quantum technologies~\cite{Fabre_2020,Brecht_2015,Mukamel_2020,Raymer_Walmsley_2020}. In spectroscopy, for example, resolving spectral lines is essential for probing the atomic and molecular properties of matter. Frequency measurements are also crucial for quantum metrology and sensing~\cite{Giovannetti_2011,Degen_2017}, as they largely rely on measuring the transition between quantised energy levels. However, the precision of these optical measurements has long been constrained by instrumental limitations. Every spectral feature has an intrinsic linewidth set by the Fourier limit—a lower limit being the inverse of the measurement time or the signal's temporal duration. The resolving power of conventional spectrometers is compounded by the Rayleigh criterion~\cite{Rayleigh_1879}, which defines the minimal resolvable separation between two spectral lines. As the separation shrinks, the uncertainty in their resolution escalates rapidly. This phenomenon, often termed Rayleigh's curse, poses a critical barrier, particularly when probing weak, photon-limited signals.

The ultimate precision in resolving spectral features is bounded by the quantum mechanical nature of the optical field. From quantum estimation theory, Tsang, Nair, and Lu~\cite{Tsang_2016} have shown that Rayleigh's criterion is not a fundamental limit, but rather an artefact of the direct intensity (DI) measurement strategies. Instead, coherent measurements can resolve arbitrarily small separations with constant, finite precision. To circumvent the classical resolution limit, implementing a coherent mode filter to select the optimal mode basis before detection, instead of DI measurements, is essential~\cite{Pa_r_2016,Rehacek_2017,Tham_2017,Gefen_2019,Pushkina_2021,Zanforlin_2022}. However, the practical realisation of this advantage relies critically on the accuracy of mode filtering, as experimental imperfections like mode crosstalk and detector noise substantially diminish the achievable precision~\cite{Bonsma_Fisher_2019,Gessner_2020,Lupo_2020,Sorelli_2021,Oh_2021,Rouvi_re_2024}. The high-fidelity coherent mode filtering is generally needed for broader quantum information processing and quantum metrology technologies~\cite{Brecht_2015,Raymer_2020,Humphreys_2014,Awschalom_2021}. Looking ahead, future quantum networks will demand sensor nodes capable of coherent TF processing, ideally integrated with on-demand buffering for signal synchronization~\cite{Makino_2016,Davidson_2023}, coherent bandwidth and frequency conversion to interface different physical platforms and network channels~\cite{Kielpinski_2011,Radnaev_2010,Karpi_ski_2016}—all while preserving quantum coherence. These integrated functionalities are essential for building robust, distributed quantum-enhanced sensors~\cite{Zhang_2021} capable of dynamically adjusting to varying environmental conditions.

In TF super-resolution, quantum pulse gates (QPGs), which use nonlinear waveguides for mode-selective frequency conversion, have been deployed to resolve temporal and spectral separations for ultrafast pulses with hundreds-of-GHz bandwidth~\cite{Donohue_2018,Ansari_2021}. While QPGs demonstrate programmable mode selectivity~\cite{Serino_2025}, they inherently lack on-demand storage and buffering capabilities. In the narrow-band frequency regime, time-inversion interferometry using gradient echo memory has demonstrated sub-Rayleigh resolution for tens-of-kHz bandwidth pulses~\cite{Mazelanik_2022}, but it is limited to a fixed symmetric-antisymmetric mode selectivity and the ultra-narrowband regime, requiring cumbersome magnetic field gradients and cryogenics to map frequency components to longitudinal spatial positions. In addition, there have been works focusing on frequency super-resolution across tens-of-GHz bandwidth using dispersion engineering and electro-optic modulators (EOM)~\cite{Shah_2021,Lipka_2024}. However, no existing platform delivers high‐precision super‐resolution in the MHz-to-GHz bandwidth together with on‐demand storage, retrieval and user-defined mode selectivity. 

Here, we introduce a TF super-resolution scheme based on high-fidelity coherent mode filtering in an atomic Raman quantum memory. Photonic quantum memories have been widely studied for absorbing and re-emitting photonic states on demand, with applications in quantum networks, communication and computing~\cite{Lvovsky_2009}. Implemented in warm caesium vapour, our platform utilises a stimulated Raman process where a strong control field with a tailored temporal profile coherently maps an incoming signal field onto a collective atomic spin-wave coherence, achieving mode selectivity up to 99.6\% for orthogonal Hermite-Gaussian (HG) temporal modes. Leveraging this user-defined coherent mode filter, we store the optimal signal mode containing the information of spectral line separation, retrieve it on demand, and apply maximum-likelihood estimation to the retrieved photon statistics to extract frequency separations. Focusing on sub-linewidth frequency separation under various detected photon budgets (from $2\times10^3$ to $1\times10^5$), our platform consistently outperforms DI methods, achieving high-precision enhancements. Operating in the MHz-to-GHz bandwidth, our memory-based platform extends the toolbox of TF metrology and provides integrated functionalities encompassing mode filtering, buffering, and shaping. These capabilities are ideally suited for next-generation quantum sensor nodes and their deployment in quantum networks.

\section{Results}
\subsection{Super-resolving measurement in frequency domain}

\par
In sensing and metrology, resolution and precision are linked to signal bandwidth and measurement time; specifically, frequency resolution scales directly with the signal bandwidth $\sigma$. We frame the task of resolving two closely spaced spectral features as estimating their frequency separation normalised by the signal bandwidth, $\epsilon = \Delta\omega / \sigma$. Here, $\Delta\omega$ denotes the frequency difference between the two spectral lines centred at $\omega_0 \pm \Delta\omega/2$. We model the sources as two mutually incoherent emitters of equal intensity, each described by a Gaussian spectral amplitude $\psi(\omega) = (2\pi\sigma^2)^{-1/4} \exp(-\omega^2 / 4\sigma^2)$. Conventional spectroscopy measures the power spectrum,
\begin{equation}
    S(\omega|\epsilon)=\frac{1}{2}\left[\left|\psi\left(\omega-\omega_0-\frac{\epsilon\sigma}{2}\right)\right|^2+\left|\psi\left(\omega-\omega_0+\frac{\epsilon\sigma}{2}\right)\right|^2\right].
    \label{eq:intensity_profile}    
\end{equation}
The separation $\epsilon$ is then estimated by fitting this model to the measured spectrum. The amount of information about $\epsilon$ extractable from the measurement outcomes is quantified by the Fisher information (FI) per detected photon
\begin{equation}
    \mathcal{F}_{\text{DI}}(\epsilon) = \int_{-\infty}^{\infty}d\omega\frac{1}{S\left(\omega|\epsilon\right)}\left(\frac{\partial S\left(\omega|\epsilon\right)}{\partial\epsilon}\right)^2,
    \label{eq:FI_DI}
\end{equation}
which vanishes in the limit $\epsilon\to0$~\cite{Tsang_2016}. The precision, quantified by the variance of an unbiased estimator $\hat{\epsilon}$, is lower-bounded by the Cramér–Rao lower bound (CRLB) $\text{Var}(\hat{\epsilon}) \geq 1/[N\mathcal{F}(\epsilon)]$, where $N$ represents the number of detected photons. As $\epsilon\to0$, the two Gaussian lines increasingly overlap and the estimator variance diverges to infinity, making the estimation of arbitrarily small separations infeasible.

An optimal measurement, in contrast, projects the signal onto an orthonormal HG mode basis with coherent filtering. This mode basis exhibits a constant FI of $\mathcal{F}_{\text{HG}}(\epsilon)\approx 1/4$ that saturates the quantum FI—the ultimate limit of precision attainable over all possible measurements for a given input quantum state~\cite{Tsang_2016}. As the HG measurement is particularly advantageous for small separations, we will focus on the precision and ability to resolve small separations under limited photon budgets in realistic experiments. For small separations, the projection onto the $\text{HG}_0$ and $\text{HG}_1$ modes contributes most of the FI. The ideal projection probabilities onto HG$_0$ and HG$_1$ modes define a raw estimator $\hat{\epsilon}_\text{raw}=4\sqrt{N_1/N_0}$, where $N_0$ and $N_1$ are the experimentally measured counts of HG$_0$ and HG$_1$ projections.

In practical implementations, mode crosstalk modifies the ideal projection probabilities. We can model this effect between the first two HG modes by describing how the ideal probability vector, $\boldsymbol{p}$, is perturbed into the measured probability vector, $\boldsymbol{\tilde{p}}$. This transformation is defined by a coupling matrix $\boldsymbol{M}$, as
\begin{equation}
    \boldsymbol{\tilde{p}} = \boldsymbol{Mp} =
    \begin{pmatrix}
    \alpha & 1-\beta \\
    1-\alpha & \beta
    \end{pmatrix}
    \begin{pmatrix}
    P(0|\epsilon) \\
    P(1|\epsilon)    
    \end{pmatrix},
    \label{eq:crosstalk_matrix}    
\end{equation}
where $\alpha,\beta\in[0,1]$ quantify the mode filtering fidelity. The resulting perturbed probability $\boldsymbol{\tilde{p}}=\begin{pmatrix}\tilde{P}(0|\epsilon)\\\tilde{P}(1|\epsilon)\end{pmatrix}$ is then normalized. Assuming low mode crosstalk $\alpha\approx1$ and $\beta\approx1$, we derive the FI for small separations as (see Supplementary Section A for details)
\begin{equation}
    \mathcal{F}(\epsilon)\approx\frac{1}{4\left[(1-\alpha)/(\epsilon/4)^2+1\right]}.
    \label{eq:FI_2mode_crosstalk}    
\end{equation}
The FI in the presence of mode crosstalk is highly sensitive to the leakage from the HG$_0$ mode into the HG$_1$ mode, $1-\alpha$. If $(\epsilon/4)^2 \ll 1-\alpha$, the FI decreases significantly from the ideal case, highlighting the importance of low-crosstalk mode filtering in resolving very small separations.

To estimate $\epsilon$ in the presence of crosstalk, we use maximum likelihood estimation (MLE) based on the perturbed projection probabilities onto the first two HG modes, $\tilde{P}(0|\epsilon)$ and $\tilde{P}(1|\epsilon)$. The MLE finds the parameter value that maximises the likelihood function $\mathcal{L}(\boldsymbol{x}|\epsilon)=\prod_{i=1}^N \tilde{P}(x_i|\epsilon)$ for observing the measured data $\boldsymbol{x}=\{x_1,x_2,\ldots,x_N\}$. For small separations, the MLE estimator can be written as
\begin{equation}
    \hat{\epsilon}_\text{MLE} =\underset{\epsilon}{\mathrm{argmax}}[N_0\ln{\tilde{P}(0|\epsilon)}+N_1\ln{\tilde{P}(1|\epsilon)}],
    \label{eq:MLE_estimator}
\end{equation}
subject to the parameter space constraint $\Theta = \{\epsilon \in \mathbb{R} \mid \epsilon \ge 0\}$. The MLE is asymptotically unbiased and efficient in the limit $N\to\infty$, implying its variance approaches the CRLB.

In real experiments with finite statistics, the MLE exhibits a non-zero bias, particularly at small separations. This bias arises from both higher-order asymptotic terms~\cite{Hervas_2025} and the non-negativity parameter space constraint $\Theta$. Owing to shot noise, the experimentally observed normalised counts $\boldsymbol{f}=\begin{pmatrix}N_0/N\\N_1/N\end{pmatrix}$ may lie outside the physically valid range. When this occurs, the likelihood function $\mathcal{L}(\boldsymbol{x}|\epsilon)$ takes its maximum at the boundary of the parameter space constraint (that is, $\epsilon=0$), leading to a positively skewed distribution of the estimator. This skewing effect results in an unavoidable bias for $\epsilon\to 0$, with the magnitude of the bias determined by the estimator’s distribution. The bias diminishes with higher photon counts $N$, as statistical fluctuations on $N_0$ and $N_1$ decrease, making the estimator's distribution at the boundary less probable (see Supplementary Section A for simulation results).

We assess our measurement performance using the mean squared error (MSE), $\text{MSE}(\epsilon,N)=\langle(\hat{\epsilon}-\epsilon)^2\rangle$, incorporating both the variance and the squared bias of the estimator, $\text{MSE}(\epsilon,N)= \text{Var}(\hat{\epsilon}) +b(\epsilon, N)^2$. The MSE is lower bounded by the CRLB with a bias term~\cite{Cover_2005}
\begin{equation}
    \text{MSE}(\epsilon,N) \geq \frac{[1 + b'(\epsilon,N)]^2}{N\mathcal{F}(\epsilon)} + b(\epsilon,N)^2,
    \label{eq:MSE_bound}
\end{equation}
where $b'(\epsilon,N)$ represents the derivative of $b(\epsilon,N)$ with respect to $\epsilon$. Only as $b(\epsilon,N)\rightarrow0$ does the MSE converge to the standard unbiased CRLB, $\text{Var}(\hat{\epsilon}) \geq 1/[N\mathcal{F}(\epsilon)]$. From the MSE, we quantify the sensitivity of the apparatus as the minimal resolvable separation $\epsilon_\text{min}$ for which the parameter-to-error ratio (PER) of the estimate,
\begin{equation}
    \text{PER}(\epsilon,N)=\epsilon^2/\text{MSE}(\epsilon,N),
    \label{eq:PER}
\end{equation}
is greater than 1~\cite{Gessner_2020}. This minimal resolvable separation defines a modified Rayleigh criterion for resolving spectral lines under certain detected photon budgets. In this work, MSE and PER serve as the figures of merit to quantify the performance of our experimental platform.

\subsection{Mode-selective Raman quantum memory}

\begin{figure*}
    \centering
    \includegraphics[width=1\linewidth]{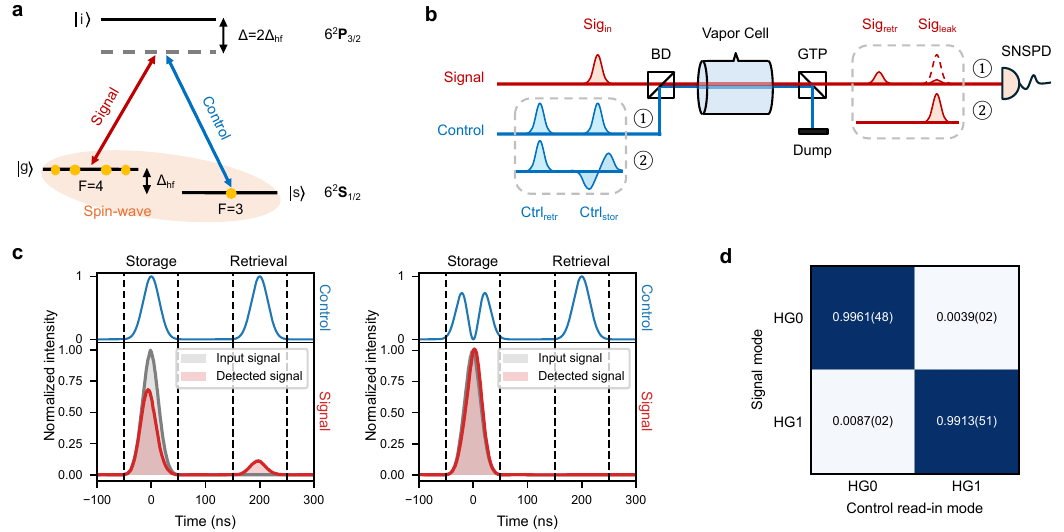}
    \caption{\textbf{Mode-selective Raman quantum memory.} (\textbf{a}) Raman memory energy levels. The states $|g\rangle$ and $|s\rangle$ are the hyperfine levels of the atomic ground state, separated by an energy splitting of $\Delta_{\text{hf}}=9.2$~GHz. Atoms are initially prepared in state $|g\rangle$. A weak signal field and a strong control field coherently map the incoming signal onto a collective atomic spin-wave coherence stored between $|g\rangle$ and $|s\rangle$. (\textbf{b}) Schematic of a Raman memory as a coherent temporal mode filter. The signal and control fields co-propagate through a vapour cell after being combined by a BD. Following interaction, the signal is detected by SNSPDs, while the control field is suppressed by a GTP and a series of etalons. When both the input signal Sig$_\text{in}$ and the control read-in pulse Ctrl$_\text{stor}$ are in the HG$_0$ mode, storage occurs (the leaked signal, Sig$_\text{leak}$, is less than Sig$_\text{in}$), followed by successful retrieval (Sig$_\text{retr}$) using the control read-out pulse (Ctrl$_\text{retr}$) (labelled as 1). In contrast, when Sig$_\text{in}$ is HG$_0$ and Ctrl$_\text{stor}$ is HG$_1$, minimal storage or retrieval is observed, demonstrating mode selectivity (labelled as 2). (\textbf{c}) Examples of experimentally detected signal sequences for the HG$_0$ signal stored using HG$_0$ and HG$_1$ control modes, respectively. (\textbf{d}) Measured mode crosstalk matrix. The matrix elements are calculated from the corresponding total efficiencies, normalised across each row. The uncertainties assume Poissonian shot noise.}
    \label{fig:memory}
\end{figure*}

\noindent
To achieve frequency-domain super-resolution, we deploy a Raman quantum memory to perform coherent temporal mode filtering. The Raman memory operates within a warm vapour ensemble of caesium-133 atoms, featuring a $\Lambda$-type three-level system as shown in Fig.~\ref{fig:memory}(a). We select the two hyperfine ground states, $|F=4\rangle$ and $|F=3\rangle$ of the $6^2\textbf{S}_{1/2}$ manifold, as the initial state $|g\rangle$ and storage states $|s\rangle$, respectively, and use the $6^2\textbf{P}_{3/2}$ level as the intermediate excited state $|i\rangle$. A strong classical control field (coupled to the $|s\rangle \leftrightarrow |i\rangle$ transition with a detuning $\Delta$) coherently maps a weak signal field (coupled to the $|g\rangle \leftrightarrow |i\rangle$ transition with the same detuning) into a collective atomic coherence (spin-wave) between $|g\rangle$ and $|s\rangle$. As shown in Fig.~\ref{fig:memory}(b), the orthogonally polarised signal and control fields are combined using a beam displacer (BD) and co-propagate through the vapour cell. The read-in control pulse Ctrl$_\text{stor}$ stores the signal pulse Sig$_\text{in}$ as spin-wave, while unabsorbed light exits as the leaked signal Sig$_\mathrm{leak}$. Retrieval is performed on demand by a second control pulse, Ctrl$_\mathrm{retr}$, which converts the stored excitation back into an optical field, Sig$_\mathrm{retr}$. By shaping the temporal profile of Ctrl$_\mathrm{retr}$, we retrieve the signal into a user-defined temporal mode, which may differ from the original read-in mode. After the memory, the control field is suppressed using a Glan–Taylor polariser (GTP) and three double-passed Fabry–Pérot etalons, leaving only the retrieved signal for detection via superconducting nanowire single-photon detectors (SNSPDs).

In the low-coupling regime (that is, at low storage efficiencies), the memory operates as a fully programmable single-mode device where the stored temporal mode is directly defined by the temporal profile of the control pulse~\cite{Nunn_2008} (see Supplementary Section B for details). Fig.~\ref{fig:memory}(b) shows two representative cases: when both signal and control are in the $\mathrm{HG}_0$ mode (labelled 1), the signal is efficiently stored and retrieved; conversely, when the signal is in $\mathrm{HG}_0$ but the control is in $\mathrm{HG}_1$ (labelled 2), storage and retrieval are strongly suppressed. Experimental results for these scenarios are shown in Fig.~\ref{fig:memory}(c). To quantify the mode selectivity, we measured the total efficiencies for all combinations of signal and control pulses prepared in HG$_0$ and HG$_1$ modes. Fig.~\ref{fig:memory}(d) shows the measured mode crosstalk matrix with high mode selectivity: $(99.61 \pm 0.48)\%$ when storing $\mathrm{HG}_0$ signals with $\mathrm{HG}_0$ control, and $(0.39 \pm 0.02)\%$ with $\mathrm{HG}_1$ control.  These measurements used read-in and read-out control pulses of $125 \pm 1$~pJ, yielding total efficiencies of about 11\% along the diagonal. At higher coupling strengths (for example, stronger control pulses), this single-mode behaviour degrades owing to increased storage of modes orthogonal to the target control mode. This highlights an inherent trade-off between mode selectivity and overall memory efficiency. By achieving low crosstalk and high mode selectivity in the low-coupling regime, our Raman memory provides a robust and effective platform for the super-resolution task.

\subsection{Experimental super-resolution with mode-selective quantum memory}

\begin{figure}
    \centering
    \includegraphics[width=1\linewidth]{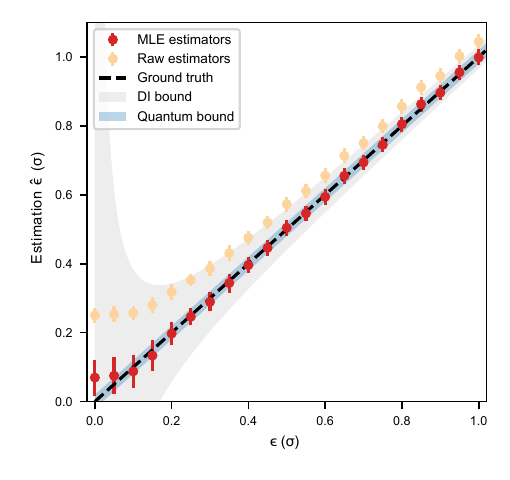}
    \caption{\textbf{Experimental estimation results.} MLEs (red) with raw estimators (yellow), both derived from $10^4$ total detected photons. The markers and error bars denote the means and standard deviations derived from 50 bootstrap resamples. The dashed black line indicates the ground truth separation. The shaded regions illustrate the theoretically predicted standard deviations from the quantum CRLB (blue) and the DI measurement CRLB (grey).}
    \label{fig:results}
\end{figure}

\noindent
To perform estimation experiments using the mode-selective Raman quantum memory described above, we prepared the signal as an incoherent mixture of two Gaussian spectral lines, each with a linewidth of $\sigma = 5.30$~MHz, and varied their normalised separation from 0 to 1 in steps of 0.05 (see Methods). For each separation, we performed two measurements by storing the signal with either an $\mathrm{HG}_0$ or an $\mathrm{HG}_1$ control read-in pulse. In both cases, the stored signals were retrieved after 200~ns using an $\mathrm{HG}_1$ pulse and detected with SNSPDs. At $\epsilon=0$, the storage and total internal efficiency for $\text{HG}_0$ read-in were $(39.5 \pm 0.6)\%$ and $(14.6 \pm 0.1)\%$ respectively, using control pulses with energies of $130 \pm 1$~pJ. For each separation, we recorded approximately $N\approx 2 \times10^5$ detector clicks in total. 

To estimate the frequency separation $\epsilon$, we first calibrated the system with $1.6\times10^5$ detected counts per separation. By fitting the experimental model to this calibration data, we determined the measured crosstalk from $\text{HG}_0$ to $\text{HG}_1$ in the estimation experiment to be 0.34\%. This calibration was performed in a separate experiment from the previous mode-filtering demonstration, accounting for the slight difference in reported crosstalk. With the calibrated model, in each run, we performed MLE to compute the estimate $\hat{\epsilon}_\text{MLE}$ based on Eq.~\ref{eq:MLE_estimator} with registered counts $(N_0,N_1)$. We then obtained the MSE for each separation by averaging $(\hat{\epsilon}_\text{MLE} - \epsilon)^2$ over 50 bootstrapped experiments with the same $N$ drawn from the full click dataset. The uncertainty in the MSE was evaluated by repeating this procedure 10 times.

Figure~\ref{fig:results} shows the estimation results with estimator standard deviations for $N=1\times10^4$ detected counts. The complete estimation results for various detection counts, detailing the contributions from estimator bias and variance, are shown in the Supplementary Section C. While the raw estimators $4\sqrt{N_1/N_0}$ have low variance, they suffer from a systematic bias caused by mode crosstalk and control field leakage. In contrast, the MLE exhibits substantially reduced bias. For the smallest separations ($\epsilon=0 - 0.2$), the MLE estimators display lower bias but higher variance, reflecting the trade-off between estimator variance (precision) and bias (accuracy) described in Eq.~\ref{eq:MSE_bound}. Crucially, all MLE error bars are smaller than the DI bound, demonstrating enhanced precision over DI detection. To validate the platform's unique functionality of on-demand storage, retrieval and coherent mode conversion, we also performed super-resolving estimation with storage times from $150$~ns to $250$~ns and with $\mathrm{HG}_0$ and $\mathrm{HG}_1$ retrieval modes. The results are included in the Supplementary Section D.

\begin{figure*}
    \centering
    \includegraphics[width=1\linewidth]{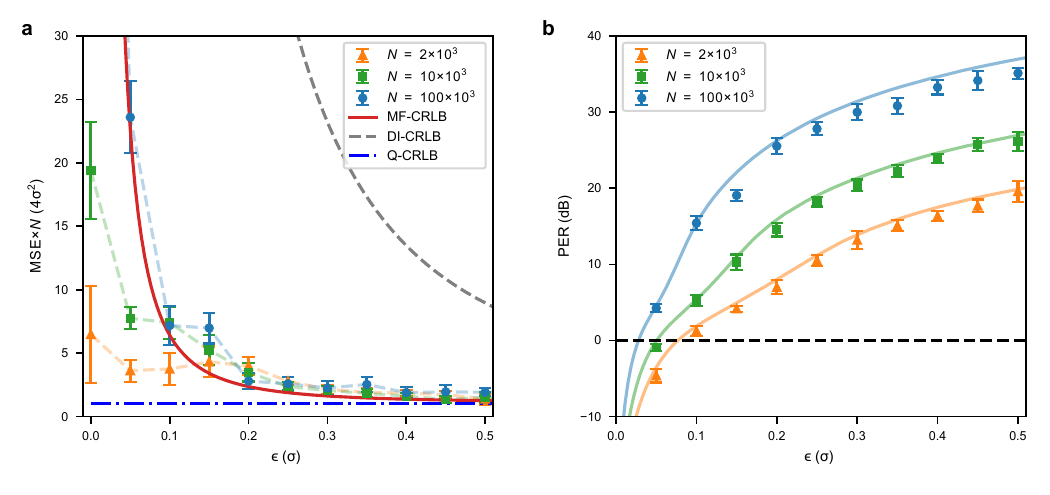}
    \caption{\textbf{Estimation performance under different photon budgets.} (\textbf{a}) MSE scaled by photon number $N$. Experimental results for $N = 2\times10^3$ (orange triangles), $10\times10^3$ (green squares), and $100\times10^3$ (blue circles) approach the theoretical CRLB of our setup (solid red line) as $N$ increases. The quantum CRLB (dash-dotted blue line) and the DI detection CRLB (dashed grey line) are shown for reference. (\textbf{b}) PER for different detection counts. Experimental data (markers) are plotted alongside their corresponding theoretical values (solid lines). The dashed line at 0 dB marks the sensitivity threshold. The markers and error bars denote the means and standard deviations derived from ten bootstrap resamples for both plots.}
    \label{fig:results_2}
\end{figure*}

The residual MLE bias near $\epsilon=0$ is mainly attributed to the non-negativity constraint on $\epsilon$. This bias is unavoidable under limited photon budgets and depends on the number of detected photons. Fig.~\ref{fig:results_2}(a) plots $\mathrm{MSE}\times N$ together with the corresponding CRLBs for comparison across different detected counts. For $\epsilon<0.5$, the measured MSE is generally smaller than the DI CRLB, indicating improved precision. However, at low photon numbers (for example, $N=2\times10^3$), the MSE falls below the mode-filtering CRLB for $\epsilon<0.15$ owing to estimation bias. As $N$ increases to $100\times10^3$, the MSE approaches the CRLB as the bias decreases, consistent with earlier discussion. Thus, while the MLE introduces a small bias at low photon numbers and small separations, its contribution to the MSE becomes negligible at larger $N$, enabling reliable high-precision estimation.

\subsection{Sensitivity and precision enhancement benchmarking}

\noindent
To benchmark the sensitivity of our measurement scheme, we examine the minimum resolvable separation using the PER defined in Eq.~\ref{eq:PER}. Fig.~\ref{fig:results_2}(b) shows the PER as a function of separation for different total detection counts. With $100\times10^3$ detected photons, we achieve $\mathrm{PER}=4.4\pm0.5$~dB at $\epsilon=0.05$, corresponding to resolving spectral lines separated by just $265$~kHz for a linewidth of $5.30$~MHz. In contrast, for lower photon numbers ($N=2\times10^3$ and $10\times10^3$), $\mathrm{PER}<0$~dB at $\epsilon=0.05$, indicating that such small separations cannot be distinguished from zero separation at these photon levels. The theoretical PER values (Fig.~\ref{fig:results_2}(b), solid lines) agree well with the experimental results.

\begin{figure*}
    \centering
    \includegraphics[width=1\linewidth]{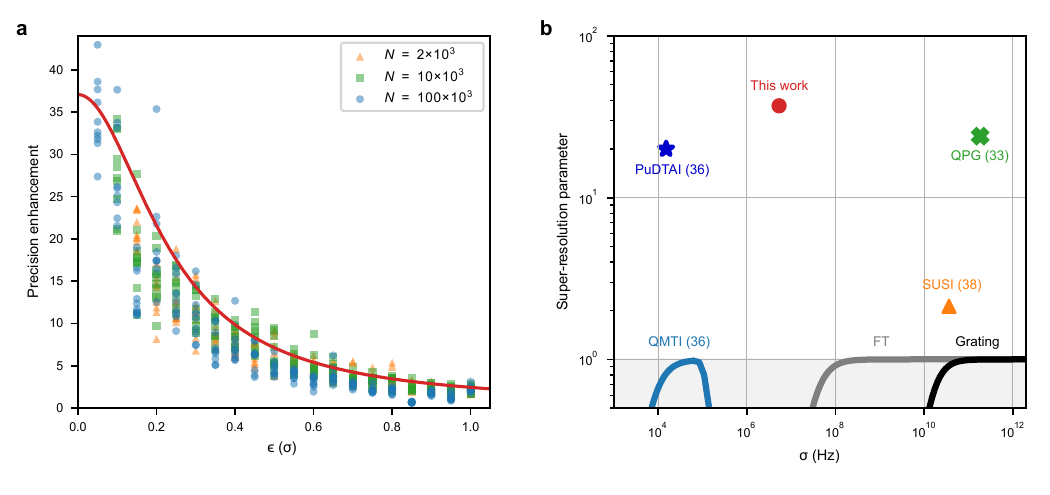}
    \caption{\textbf{Precision enhancement benchmarking.} (\textbf{a}) Precision enhancement as a function of spectral line separation $\epsilon$. The theoretical precision enhancement (red curve) is plotted with all experimental results for various photon counts: $N = 2\times10^3$ (orange triangles), $10\times10^3$ (green squares), and $100\times10^3$ (blue circles). (\textbf{b}) Comparison of super-resolution parameters for various schemes. This panel benchmarks the performance of our system against other super-resolution techniques and DI methods. QPG: quantum pulse gate~\cite{Donohue_2018}; PuDTAI: pulse division time-axis-inversion~\cite{Mazelanik_2022}; SUSI: super-resolution via spectral inversion~\cite{Lipka_2024}; QMTI: quantum-memory temporal imaging~\cite{Mazelanik_2022}; FT: Fourier transform spectrometer (Bruker IFS 125HR); Grating: a grating spectrometer of 1,200 lines per mm.}
    \label{fig:comparison}
\end{figure*}

In addition to sensitivity, we evaluate the precision enhancement enabled by our memory-based mode filtering scheme against DI methods. We quantify the enhancement via the super-resolution parameter $\mathfrak{s}$ introduced by Mazelanik et al.~\cite{Mazelanik_2022}, defined as the ratio of the FI for the two methods in the limit of $\epsilon\to 0$
\begin{equation}
    \mathfrak{s}=\lim\limits_{\epsilon\rightarrow0}(\mathcal{F}/\mathcal{F}_{\text{DI}}).
    \label{eq:super_parameter}
\end{equation}
This parameter characterises the achievable precision enhancement in the infinite photon limit, where a truly unbiased estimator exists only as $N\to\infty$. This parameter can be used to assess and compare the performance of various super-resolving measurement schemes, with its value primarily determined by mode crosstalk. Figure~\ref{fig:comparison}(a) shows how the theoretical enhancement $\mathcal{F}/\mathcal{F}_{\mathrm{DI}}$ (red curve) varies with $\epsilon$. In the limit $\epsilon \rightarrow 0$, the ratio asymptotically approaches 37 for our platform. This value sets a new benchmark for frequency-separation estimation, surpassing previous results~\cite{Donohue_2018,Mazelanik_2022,Lipka_2024}, as shown in Fig.~\ref{fig:comparison}(b). For comparison, DI methods--including quantum-memory temporal imaging~\cite{Mazelanik_2022}, Fourier transform spectroscopy and grating spectroscopy--all yield $\mathfrak{s}<1$. Among techniques spanning various bandwidth regimes, our Raman memory uniquely operates in the MHz-to-GHz range, set by the hyperfine splitting and the memory storage time. For a comprehensive comparison, see Supplementary Section E.

Under finite statistics, the experimental precision enhancement can deviate from the theoretical values owing to estimator bias. In practice, when the bias is negligible, the enhancement can be approximated by the ratio of the measured mode-filtering MSE to the CRLB of DI methods. Shown as data points in Fig.~\ref{fig:comparison}(a), these ratios follow the trend of the theoretical prediction, subject to experimental imperfections and fluctuations. However, at very small separations, the estimator bias can cause the measured MSE to fall below the CRLB, resulting in calculated ratios exceeding theoretical values. In this region, both the mode filtering method and the DI method exhibit bias, making the determination of realistic precision enhancement more complex~\cite{Bonsma_Fisher_2019}. Our experiment demonstrates a practical $(34\pm4)$-fold improvement at $\epsilon = 0.05$ and a $(28\pm6)$-fold improvement at $\epsilon = 0.1$ using $100\times10^3$ detected photons, confirming the feasibility and effectiveness of our approach. Furthermore, we expect that DI measurements, which inherently have larger estimator variances from the FI analysis, require a higher number of photons to achieve nearly unbiased estimation, whereas the optimal mode filtering maintains lower variance even in photon-limited applications.

\section{Discussion}
\par
\noindent
We have demonstrated that all-optical control of Raman interactions enables a compact, room-temperature quantum memory to function as a programmable, high-precision TF sensor across the MHz–to-GHz regime. Beyond resolving two spectral lines, our platform’s fully programmable temporal mode filtering allows us to resolve more complex spectral structures. This flexibility is essential for multi-parameter estimation, where the optimal measurement basis can be
complicated depending on the task~\cite{_eha_ek_2017,Liu_2019,Shao_2022}; such measurements could be implemented via TF mode sorting using cascaded memories or loop architectures. Overall, the combination of high-fidelity mode filtering, on-demand storage, and mode conversion establishes this platform as a versatile tool for advanced TF metrology and photon-limited sensing.

The technology developed here holds promise for diverse applications, including high-precision clock synchronization~\cite{Gosalia_2025}, spacetime positioning~\cite{Giovannetti_2001,Lamine_2008}, photon dose-limited metrology~\cite{Mukamel_2020}, and TF-encoded quantum information processing~\cite{Brecht_2015,Raymer_2020,Awschalom_2021}. For frequency metrology, our broadband, mode-selective measurement enables high-precision sensing in regimes where traditional spectroscopic techniques become impractical. In Doppler LiDAR, which infers target velocity from frequency shifts of backscattered light, conventional techniques fall into two classes: heterodyne detection, which imposes stringent phase coherence requirements, and incoherent edge techniques using narrowband filters, which suffer from a trade-off between sensitivity and dynamic range. Our platform bypasses these bottlenecks by providing programmable, high-resolution measurements of frequency shifts. Our current demonstration (5.30 MHz bandwidth at 852 nm) resolves frequency shifts down to 265 kHz, which corresponds to a velocity resolution of 0.11 m/s and a range resolution of 38 cm at MHz measurement rates, even under low signal-to-noise ratio conditions. Furthermore, the platform’s flexibility permits complex tasks such as resolving multiple proximal targets or the joint estimation of velocity and range~\cite{Kruse_2023,Huang_2021} (see Supplementary Section F for more discussions).

Further improvements in precision require reducing mode crosstalk (as indicated in Eq.~\ref{eq:FI_2mode_crosstalk}), which primarily arises from the storage process. A key limitation is the trade-off between crosstalk and storage efficiency, as higher efficiency leads to increased crosstalk (see Supplementary Section G for discussions and simulations). Recent advances, such as the efficiency enhancement via light-matter interference (EEVI) protocol~\cite{Burdekin_2025}, offer a route to high-efficiency storage while preserving mode selectivity. Cavity-enhanced Raman memories offer a similar solution~\cite{Saunders_2016}, albeit with bandwidth constraints set by cavity finesse. Alternatively, optimal control techniques can tailor pulse shapes to simultaneously maximise storage efficiency and suppress crosstalk~\cite{Gorshkov_2008,Guo_2019}. Moreover, the practical advantage of super-resolving measurements, quantified by FI per input photon, is currently limited by system photon loss. Similar to other TF super-resolution platforms, our system exhibits a low end-to-end efficiency of about 0.3\%, despite an internal efficiency of $(14.6\pm0.1)\%$. This drop is primarily due to the stringent filtering required to suppress the co-propagating control field, which is detuned by just 9.2 GHz. Achieving the necessary $4 \times 10^7$ extinction ratio requires a series of lossy optics. In contrast, ladder-type quantum memories, such as fast-ladder memory~\cite{Davidson_2023_FLAME} and off-resonant cascaded absorption memory (ORCA)~\cite{Kaczmarek_2018,Thomas_2023}, exploit a higher-lying excited state alongside widely separated, counter-propagating signal and control fields. This configuration enables efficient control suppression with minimal loss. These systems have demonstrated noise levels as low as $10^{-5}$ photons per signal photon~\cite{Kaczmarek_2018} and end-to-end efficiencies up to 35\%~\cite{Davidson_2023_FLAME}. Integrating our high-fidelity mode-selective measurement scheme with such architectures could enable unconditional TF super-resolution.

\section{Methods}

\subsection*{Experimental setup}

\begin{figure*}
    \centering
    \includegraphics[width=1\linewidth]{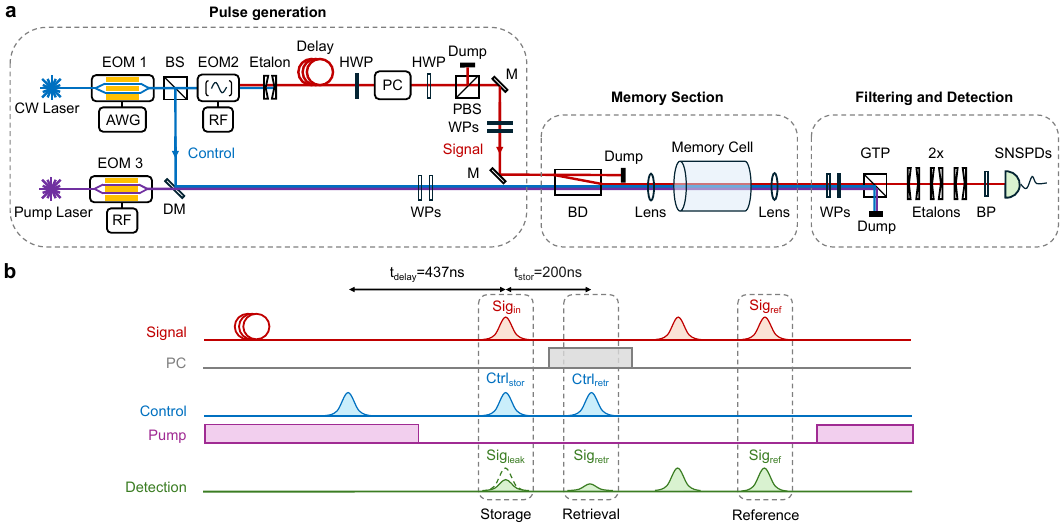}
    \caption{\textbf{Schematic of experimental set-up.} (\textbf{a}) Experimental set-up. The apparatus consists of three main sections: pulse generation, memory interaction, and filtering and detection. Both signal and control pulses originate from a CW laser at the control-field frequency. EOM 1 first carves the CW light into pulses. A 90:10 beam splitter (BS) then splits the laser into a high-power control path (90\%) and a low-power signal path (10\%). The laser in the signal path is later shifted to the signal frequency via EOM 2. The two beams are recombined via a BD before entering the memory cell for storage and retrieval. Post-retrieval filtering comprises a GTP, three double-passed etalons, and a band-pass filter (BP). The signal is finally routed to SNSPDs for detection. (PC: Pockels cell; PBS: polarising beamsplitter; M: mirror; DM: dichroic mirror; HWP: half-wave plate; WPs: a half-wave plate and a quarter-wave plate.) (\textbf{b}) Pulse sequences. The diagram illustrates the timing of the signal and control pulses, along with the gating windows of the optical pumping and the Pockels cell.}
    \label{fig:setup}
\end{figure*}

\noindent
Our Raman memory setup is shown in Fig.~\ref{fig:setup}(a). The storage medium consists of a 75~mm-long caesium (Cs-133) vapour cell with a three-layer $\mu$-metal magnetic shielding. We heated the cell to $105\pm1$~°C using a quad-twisted cryogenic wire, achieving an effective optical depth of $(4.8\pm0.1)\times10^3$. The atoms were initialised to the initial state $|g\rangle$ by an external cavity diode laser (ECDL, Toptica DL100) driving the D1 transition ($6^2\textbf{S}_{1/2}\rightarrow6^2\textbf{P}_{1/2}$). EOM 3 (EOSpace 850UL) gated the pump laser, switching it off during the experimental window. We achieved a pumping efficiency of $(99.6\pm0.1)$\%. 

We operated our memory with a detuning $\Delta=2\Delta_\text{hf}$ to suppress four-wave-mixing noise~\cite{Thomas_2019}. Both the signal and control fields were derived from a single continuous wave (CW) ECDL laser (Toptica DL Pro) operating at the control frequency (351.7122~THz). An arbitrary waveform generator (AWG, Tektronix AWG70001A) drove the EOM 1 (Sacher Lasertechnik AM830PF) to carve the CW laser into pulses. The laser was then split using a 90:10 beam splitter, directing 90\% of the optical power to the control field path. The remaining 10\% went to EOM 2 (New Focus 4851), which generated a sideband at the signal frequency (351.7030~THz). We selected this sideband as the signal field by using a Fabry-P\'erot etalon with a free spectral range (FSR) of 36.8~GHz. The signal field was delayed by 437~ns using a long fibre to achieve temporal synchronisation with the control pulses, ensuring their temporal overlap for memory storage and retrieval. Before entering the memory, the signal field also passed through a Pockels cell to remove residual CW background in the retrieval window. The polarisation of the signal and control fields was set to orthogonal for memory interaction, allowing their combination using a BD before being focused into the vapour cell.

We used a pair of convex lenses to focus the signal, control and pumping beams into the vapour cell to enhance the interaction strength. The beam widths at the focus were approximately $164\pm5$~$\mu$m for the signal and $190\pm5$~$\mu$m for the control. As shown in Fig.~\ref{fig:setup}(b), the first signal pulse temporally overlaps with the control pulse Ctrl$_\text{stor}$ for storage. After a storage time of 200~ns, the control pulse Ctrl$_\text{retr}$ retrieves the stored signal. The third signal pulse serves as a reference for the signal pulse power. All signal and control pulses share the same Gaussian envelope width parameter as a fundamental Gaussian pulse (HG$_0$) with an electric field full width at half maximum of 50~ns.

After the memory, we suppressed the control and pump fields using a GTP, followed by three double-passed etalons (two with FSR of 18.4~GHz and one with FSR 103~GHz) and a bandpass filter (central wavelength: 850~nm, full width at half maximum bandwidth: 10~nm). Finally, we split the signal into four equal parts and detected them using four SNSPDs (Photon Spot). Using multiple detectors helps mitigate detector dead time effects and enables higher overall count rates. The detected counts were registered using a time tagger (Swabian TimeTagger20).

\subsection*{Signal preparation}

\noindent
The signal in our separation estimation task is an incoherent mixture of two Gaussian spectral lines. Its frequency-domain representation can be written as
\begin{equation}
    \begin{aligned}
    \Psi(\omega)= \frac{1}{\sqrt{2}} & \left[  \exp\left({-i\frac{\phi}{2}}\right)  \right. \psi(\omega-\omega_0-\frac{\epsilon\sigma}{2})  \\
        & +\left. \exp\left({i\frac{\phi}{2}}\right)\psi(\omega-\omega_0+\frac{\epsilon\sigma}{2}) \right],
    \end{aligned}
\label{eqn:SPE_modulatedsignal_freq}
\end{equation}
where $\phi \in (-\pi,\pi]$ is a random phase between the two spectral lines. Here, we carved the temporal profile of our signal field as the inverse Fourier transform of Eq.~\ref{eqn:SPE_modulatedsignal_freq},
\begin{equation}
    \Psi(t) = A\cos\left(\frac{\epsilon\sigma t-\phi}{2}\right)\exp(-t^2\sigma^2),
    \label{eq:Signal_temporal}
\end{equation}
where $A$ is the amplitude. The signal pulse has a spectral width of $\sigma = 33.3$~Mrad/s (5.30~MHz). In our experiment, we generated signals with separation parameters $\epsilon$ from 0 to 1 with an increment of 0.05 (0.265~MHz). To introduce the incoherence of the two spectral lines, we prepared the signal in a mixture of four different relative phases, $\phi=-\pi/2,0,\pi/2,\pi$, for each separation.

One key requirement for this experiment is the accurate generation of signal and control pulses. To ensure high-fidelity pulse carving, we characterised the frequency response of our electronic pulse generation system, which comprises the AWG, a radio-frequency (RF) splitter, and two RF amplifiers that drive the EOM 1. By measuring the system's frequency response function, we applied a correction to the input signals of the AWG, compensating for frequency-dependent variations in the RF components. This correction ensures uniform amplification across all frequency components, minimising distortions in the carved pulses. The measured intensity pulse shapes of HG$_0$ and HG$_1$ are presented in the Supplementary Section H.

\subsection*{Data collection and analysis}

\noindent
For each signal, comprising one separation and one phase setting, we first stored it using a control pulse of HG$_0$ with a pulse energy of $130\pm1$~pJ.  At this control energy, our memory typically operates with a storage efficiency of $(39.5\pm0.6)\%$ and a total internal efficiency of $(14.6\pm0.1)\%$ for a storage duration of 200~ns. The stored signal was retrieved using a control pulse in the HG$_1$ mode, with the same pulse energy as the HG$_0$ write-in pulse. The HG$_1$ mode was chosen to minimise distortion in the pulse sequence due to its symmetrical temporal profile. Importantly, the choice of retrieval mode has a negligible impact on the retrieval efficiency. The pulse sequence was repeated every 3~$\mu$s, with a total measurement time of 2~s. To accurately account for signal background noise and dark counts, we also recorded data with the control pulse blocked for an equivalent duration. The counts detected in the retrieval time window during this control-blocked run were later used to subtract the noise counts from the retrieved signal counts to minimise the estimator bias caused by these noises. The above procedures were repeated for all four phases. Subsequently, we modified the control read-in pulse to the HG$_1$ mode while keeping the read-out pulse unchanged and repeated the measurements. Finally, the entire experiment was repeated for all separations of interest.

To obtain the final detected counts, we first identified a reference pulse count that yielded approximately $N$ = $2\times10^3$, $10\times10^3$, or $100\times10^3$ total retrieval counts (with a variation less than 5\%) of the signals across all combinations of the two control read-in pulses and four phases. For each separation, we used the reference counts to randomly sample a subset that yielded $N$ retrieved counts from the data files. The corresponding noise counts were subtracted from the retrieved counts to correct the background contributions. This random sampling approach, combined with the use of reference counts, was used to mitigate the impact of signal power fluctuations on the final estimation process. The retrieval counts acquired for HG$_0$ and HG$_1$ storage, $N_0$ and $N_1$ respectively, were used directly to compute the raw estimator. We repeated the entire sampling and analysis procedure 50 times using bootstrapping to estimate the variances and MSEs of the estimators. To obtain the error bars for the MSE shown in Fig.~\ref{fig:results_2}, we repeated the bootstrapping process 10 times to get 10 datasets and calculated the standard deviations of the MSE.

\subsection*{System calibration}

\noindent
To characterize the crosstalk matrix $\boldsymbol{M}$ of the experimental system and the perturbed HG projection probabilities $\tilde{P}(0|\epsilon)$ and $\tilde{P}(1|\epsilon)$ in Eq.~\ref{eq:MLE_estimator}, we used a dataset for separations $\epsilon$ from 0 to 1 with $N=1.6\times10^5$ detected photon counts for each separation. For each separation, we normalised the retrieved counts $N_0$ and $N_1$ corresponding to the two HG projections to obtain the relative frequencies $f_0=N_0/(N_0+N_1)$ and $f_1=N_1/(N_0+N_1)$. The values of $\alpha$ and $\beta$ were estimated by minimising the least square cost

\begin{equation*}
\begin{split}
C( & f_0,f_1| \alpha,\beta) \\
& = \sum_{\epsilon}\left\{\left[f_0(\epsilon)-\tilde{P}(0|\epsilon)\right]^2 + \left[f_1(\epsilon) - \tilde{P}(1|\epsilon)\right]^2\right\}.
\end{split}
\end{equation*}
The perturbed probabilities, parameterised by $\alpha$ and $\beta$, were compared to the measured relative frequencies, and the sum of squared residuals across all separations was computed. We then optimised the parameters using nonlinear least-squares minimisation. The resulting values characterise the performance of our experimental system and were subsequently used to compute the MLE estimates.

\section*{Acknowledgments}
We acknowledge J. Szuniewicz for help in developing the EOM control code and R. B. Patel for help in setting up the detectors. We thank B. Brecht, M. G. Raymer, A.M. Steinberg and S. Yu for helpful discussions. This work was supported by the European Union’s Horizon 2020 Research and Innovation Programme Grant No. 899587 Stormytune and the Engineering and Physical Sciences Research Council via the Quantum Computing and Simulation Hub (Grant No. T001062). A. Z. acknowledges a UK Research and Innovation Guarantee Postdoctoral Fellowship under the UK government’s Horizon Europe funding Guarantee (EP/Y029127/1). S.E.T. acknowledges an Imperial College Research Fellowship.


\begin{thebibliography}{61}%
\makeatletter
\providecommand \@ifxundefined [1]{%
 \@ifx{#1\undefined}
}%
\providecommand \@ifnum [1]{%
 \ifnum #1\expandafter \@firstoftwo
 \else \expandafter \@secondoftwo
 \fi
}%
\providecommand \@ifx [1]{%
 \ifx #1\expandafter \@firstoftwo
 \else \expandafter \@secondoftwo
 \fi
}%
\providecommand \natexlab [1]{#1}%
\providecommand \enquote  [1]{``#1''}%
\providecommand \bibnamefont  [1]{#1}%
\providecommand \bibfnamefont [1]{#1}%
\providecommand \citenamefont [1]{#1}%
\providecommand \@href[1]{\@@startlink{#1}\@@href}%
\providecommand \@@href[1]{\endgroup#1\@@endlink}%
\providecommand \@sanitize@url [0]{\catcode `\\12\catcode `\$12\catcode `\&12\catcode `\#12\catcode `\^12\catcode `\_12\catcode `\%12\relax}%
\providecommand \@@startlink[1]{}%
\providecommand \@@endlink[0]{}%
\providecommand \@url [1]{\endgroup\@href {#1}{\urlprefix }}%
\providecommand \urlprefix  [0]{URL }%
\providecommand \doibase [0]{https://doi.org/}%
\providecommand \selectlanguage [0]{\@gobble}%
\providecommand \bibinfo  [0]{\@secondoftwo}%
\providecommand \bibfield  [0]{\@secondoftwo}%
\providecommand \translation [1]{[#1]}%
\providecommand \BibitemOpen [0]{}%
\providecommand \bibitemStop [0]{}%
\providecommand \bibitemNoStop [0]{.\EOS\space}%
\providecommand \EOS [0]{\spacefactor3000\relax}%
\providecommand \BibitemShut  [1]{\csname bibitem#1\endcsname}%
\let\auto@bib@innerbib\@empty
\bibitem [{\citenamefont {Udem}\ \emph {et~al.}(2002)\citenamefont {Udem}, \citenamefont {Holzwarth},\ and\ \citenamefont {Hänsch}}]{Udem_2002}%
  \BibitemOpen
  \bibfield  {author} {\bibinfo {author} {\bibfnamefont {T.}~\bibnamefont {Udem}}, \bibinfo {author} {\bibfnamefont {R.}~\bibnamefont {Holzwarth}},\ and\ \bibinfo {author} {\bibfnamefont {T.~W.}\ \bibnamefont {Hänsch}},\ }\bibfield  {title} {\bibinfo {title} {Optical frequency metrology},\ }\href {https://doi.org/10.1038/416233a} {\bibfield  {journal} {\bibinfo  {journal} {Nature}\ }\textbf {\bibinfo {volume} {416}},\ \bibinfo {pages} {233} (\bibinfo {year} {2002})}\BibitemShut {NoStop}%
\bibitem [{\citenamefont {Ludlow}\ \emph {et~al.}(2015)\citenamefont {Ludlow}, \citenamefont {Boyd}, \citenamefont {Ye}, \citenamefont {Peik},\ and\ \citenamefont {Schmidt}}]{Ludlow_2015}%
  \BibitemOpen
  \bibfield  {author} {\bibinfo {author} {\bibfnamefont {A.~D.}\ \bibnamefont {Ludlow}}, \bibinfo {author} {\bibfnamefont {M.~M.}\ \bibnamefont {Boyd}}, \bibinfo {author} {\bibfnamefont {J.}~\bibnamefont {Ye}}, \bibinfo {author} {\bibfnamefont {E.}~\bibnamefont {Peik}},\ and\ \bibinfo {author} {\bibfnamefont {P.~O.}\ \bibnamefont {Schmidt}},\ }\bibfield  {title} {\bibinfo {title} {Optical atomic clocks},\ }\href {https://doi.org/10.1103/RevModPhys.87.637} {\bibfield  {journal} {\bibinfo  {journal} {Reviews of Modern Physics}\ }\textbf {\bibinfo {volume} {87}},\ \bibinfo {pages} {637} (\bibinfo {year} {2015})}\BibitemShut {NoStop}%
\bibitem [{\citenamefont {Walmsley}\ and\ \citenamefont {Dorrer}(2009)}]{Walmsley:09}%
  \BibitemOpen
  \bibfield  {author} {\bibinfo {author} {\bibfnamefont {I.~A.}\ \bibnamefont {Walmsley}}\ and\ \bibinfo {author} {\bibfnamefont {C.}~\bibnamefont {Dorrer}},\ }\bibfield  {title} {\bibinfo {title} {Characterization of ultrashort electromagnetic pulses},\ }\href {https://doi.org/10.1364/AOP.1.000308} {\bibfield  {journal} {\bibinfo  {journal} {Adv. Opt. Photon.}\ }\textbf {\bibinfo {volume} {1}},\ \bibinfo {pages} {308} (\bibinfo {year} {2009})}\BibitemShut {NoStop}%
\bibitem [{\citenamefont {Fabre}\ and\ \citenamefont {Treps}(2020)}]{Fabre_2020}%
  \BibitemOpen
  \bibfield  {author} {\bibinfo {author} {\bibfnamefont {C.}~\bibnamefont {Fabre}}\ and\ \bibinfo {author} {\bibfnamefont {N.}~\bibnamefont {Treps}},\ }\bibfield  {title} {\bibinfo {title} {Modes and states in quantum optics},\ }\href {https://doi.org/10.1103/RevModPhys.92.035005} {\bibfield  {journal} {\bibinfo  {journal} {Reviews of Modern Physics}\ }\textbf {\bibinfo {volume} {92}},\ \bibinfo {pages} {035005} (\bibinfo {year} {2020})}\BibitemShut {NoStop}%
\bibitem [{\citenamefont {Brecht}\ \emph {et~al.}(2015)\citenamefont {Brecht}, \citenamefont {Reddy}, \citenamefont {Silberhorn},\ and\ \citenamefont {Raymer}}]{Brecht_2015}%
  \BibitemOpen
  \bibfield  {author} {\bibinfo {author} {\bibfnamefont {B.}~\bibnamefont {Brecht}}, \bibinfo {author} {\bibfnamefont {D.~V.}\ \bibnamefont {Reddy}}, \bibinfo {author} {\bibfnamefont {C.}~\bibnamefont {Silberhorn}},\ and\ \bibinfo {author} {\bibfnamefont {M.~G.}\ \bibnamefont {Raymer}},\ }\bibfield  {title} {\bibinfo {title} {Photon temporal modes: A complete framework for quantum information science},\ }\href {https://doi.org/10.1103/PhysRevX.5.041017} {\bibfield  {journal} {\bibinfo  {journal} {Physical Review X}\ }\textbf {\bibinfo {volume} {5}},\ \bibinfo {pages} {041017} (\bibinfo {year} {2015})}\BibitemShut {NoStop}%
\bibitem [{\citenamefont {Mukamel}\ \emph {et~al.}(2020)\citenamefont {Mukamel}, \citenamefont {Freyberger}, \citenamefont {Schleich}, \citenamefont {Bellini}, \citenamefont {Zavatta}, \citenamefont {Leuchs}, \citenamefont {Silberhorn}, \citenamefont {Boyd}, \citenamefont {Sánchez-Soto}, \citenamefont {Stefanov}, \citenamefont {Barbieri}, \citenamefont {Paterova}, \citenamefont {Krivitsky}, \citenamefont {Shwartz}, \citenamefont {Tamasaku}, \citenamefont {Dorfman}, \citenamefont {Schlawin}, \citenamefont {Sandoghdar}, \citenamefont {Raymer}, \citenamefont {Marcus}, \citenamefont {Varnavski}, \citenamefont {Goodson}, \citenamefont {Zhou}, \citenamefont {Shi}, \citenamefont {Asban}, \citenamefont {Scully}, \citenamefont {Agarwal}, \citenamefont {Peng}, \citenamefont {Sokolov}, \citenamefont {Zhang}, \citenamefont {Zubairy}, \citenamefont {Vartanyants}, \citenamefont {del Valle},\ and\ \citenamefont {Laussy}}]{Mukamel_2020}%
  \BibitemOpen
  \bibfield  {author} {\bibinfo {author} {\bibfnamefont {S.}~\bibnamefont {Mukamel}}, \bibinfo {author} {\bibfnamefont {M.}~\bibnamefont {Freyberger}}, \bibinfo {author} {\bibfnamefont {W.}~\bibnamefont {Schleich}}, \bibinfo {author} {\bibfnamefont {M.}~\bibnamefont {Bellini}}, \bibinfo {author} {\bibfnamefont {A.}~\bibnamefont {Zavatta}}, \bibinfo {author} {\bibfnamefont {G.}~\bibnamefont {Leuchs}}, \bibinfo {author} {\bibfnamefont {C.}~\bibnamefont {Silberhorn}}, \bibinfo {author} {\bibfnamefont {R.~W.}\ \bibnamefont {Boyd}}, \bibinfo {author} {\bibfnamefont {L.~L.}\ \bibnamefont {Sánchez-Soto}}, \bibinfo {author} {\bibfnamefont {A.}~\bibnamefont {Stefanov}}, \bibinfo {author} {\bibfnamefont {M.}~\bibnamefont {Barbieri}}, \bibinfo {author} {\bibfnamefont {A.}~\bibnamefont {Paterova}}, \bibinfo {author} {\bibfnamefont {L.}~\bibnamefont {Krivitsky}}, \bibinfo {author} {\bibfnamefont {S.}~\bibnamefont {Shwartz}}, \bibinfo {author} {\bibfnamefont {K.}~\bibnamefont {Tamasaku}}, \bibinfo {author} {\bibfnamefont
  {K.}~\bibnamefont {Dorfman}}, \bibinfo {author} {\bibfnamefont {F.}~\bibnamefont {Schlawin}}, \bibinfo {author} {\bibfnamefont {V.}~\bibnamefont {Sandoghdar}}, \bibinfo {author} {\bibfnamefont {M.}~\bibnamefont {Raymer}}, \bibinfo {author} {\bibfnamefont {A.}~\bibnamefont {Marcus}}, \bibinfo {author} {\bibfnamefont {O.}~\bibnamefont {Varnavski}}, \bibinfo {author} {\bibfnamefont {T.}~\bibnamefont {Goodson}}, \bibinfo {author} {\bibfnamefont {Z.-Y.}\ \bibnamefont {Zhou}}, \bibinfo {author} {\bibfnamefont {B.-S.}\ \bibnamefont {Shi}}, \bibinfo {author} {\bibfnamefont {S.}~\bibnamefont {Asban}}, \bibinfo {author} {\bibfnamefont {M.}~\bibnamefont {Scully}}, \bibinfo {author} {\bibfnamefont {G.}~\bibnamefont {Agarwal}}, \bibinfo {author} {\bibfnamefont {T.}~\bibnamefont {Peng}}, \bibinfo {author} {\bibfnamefont {A.~V.}\ \bibnamefont {Sokolov}}, \bibinfo {author} {\bibfnamefont {Z.-D.}\ \bibnamefont {Zhang}}, \bibinfo {author} {\bibfnamefont {M.~S.}\ \bibnamefont {Zubairy}}, \bibinfo {author} {\bibfnamefont
  {I.~A.}\ \bibnamefont {Vartanyants}}, \bibinfo {author} {\bibfnamefont {E.}~\bibnamefont {del Valle}},\ and\ \bibinfo {author} {\bibfnamefont {F.}~\bibnamefont {Laussy}},\ }\bibfield  {title} {\bibinfo {title} {Roadmap on quantum light spectroscopy},\ }\href {https://doi.org/10.1088/1361-6455/ab69a8} {\bibfield  {journal} {\bibinfo  {journal} {Journal of Physics B: Atomic, Molecular and Optical Physics}\ }\textbf {\bibinfo {volume} {53}},\ \bibinfo {pages} {072002} (\bibinfo {year} {2020})}\BibitemShut {NoStop}%
\bibitem [{\citenamefont {Raymer}\ and\ \citenamefont {Walmsley}(2020)}]{Raymer_Walmsley_2020}%
  \BibitemOpen
  \bibfield  {author} {\bibinfo {author} {\bibfnamefont {M.~G.}\ \bibnamefont {Raymer}}\ and\ \bibinfo {author} {\bibfnamefont {I.~A.}\ \bibnamefont {Walmsley}},\ }\bibfield  {title} {\bibinfo {title} {Temporal modes in quantum optics: then and now},\ }\href {https://doi.org/10.1088/1402-4896/ab6153} {\bibfield  {journal} {\bibinfo  {journal} {Physica Scripta}\ }\textbf {\bibinfo {volume} {95}},\ \bibinfo {pages} {064002} (\bibinfo {year} {2020})}\BibitemShut {NoStop}%
\bibitem [{\citenamefont {Giovannetti}\ \emph {et~al.}(2011)\citenamefont {Giovannetti}, \citenamefont {Lloyd},\ and\ \citenamefont {Maccone}}]{Giovannetti_2011}%
  \BibitemOpen
  \bibfield  {author} {\bibinfo {author} {\bibfnamefont {V.}~\bibnamefont {Giovannetti}}, \bibinfo {author} {\bibfnamefont {S.}~\bibnamefont {Lloyd}},\ and\ \bibinfo {author} {\bibfnamefont {L.}~\bibnamefont {Maccone}},\ }\bibfield  {title} {\bibinfo {title} {Advances in quantum metrology},\ }\href {https://doi.org/10.1038/nphoton.2011.35} {\bibfield  {journal} {\bibinfo  {journal} {Nature Photonics}\ }\textbf {\bibinfo {volume} {5}},\ \bibinfo {pages} {222} (\bibinfo {year} {2011})}\BibitemShut {NoStop}%
\bibitem [{\citenamefont {Degen}\ \emph {et~al.}(2017)\citenamefont {Degen}, \citenamefont {Reinhard},\ and\ \citenamefont {Cappellaro}}]{Degen_2017}%
  \BibitemOpen
  \bibfield  {author} {\bibinfo {author} {\bibfnamefont {C.}~\bibnamefont {Degen}}, \bibinfo {author} {\bibfnamefont {F.}~\bibnamefont {Reinhard}},\ and\ \bibinfo {author} {\bibfnamefont {P.}~\bibnamefont {Cappellaro}},\ }\bibfield  {title} {\bibinfo {title} {Quantum sensing},\ }\href {https://doi.org/10.1103/RevModPhys.89.035002} {\bibfield  {journal} {\bibinfo  {journal} {Reviews of Modern Physics}\ }\textbf {\bibinfo {volume} {89}},\ \bibinfo {pages} {035002} (\bibinfo {year} {2017})}\BibitemShut {NoStop}%
\bibitem [{\citenamefont {Rayleigh}(1879)}]{Rayleigh_1879}%
  \BibitemOpen
  \bibfield  {author} {\bibinfo {author} {\bibnamefont {Rayleigh}},\ }\bibfield  {title} {\bibinfo {title} {Xxxi. investigations in optics, with special reference to the spectroscope},\ }\href {https://doi.org/10.1080/14786447908639684} {\bibfield  {journal} {\bibinfo  {journal} {The London, Edinburgh, and Dublin Philosophical Magazine and Journal of Science}\ }\textbf {\bibinfo {volume} {8}},\ \bibinfo {pages} {261} (\bibinfo {year} {1879})}\BibitemShut {NoStop}%
\bibitem [{\citenamefont {Tsang}\ \emph {et~al.}(2016)\citenamefont {Tsang}, \citenamefont {Nair},\ and\ \citenamefont {Lu}}]{Tsang_2016}%
  \BibitemOpen
  \bibfield  {author} {\bibinfo {author} {\bibfnamefont {M.}~\bibnamefont {Tsang}}, \bibinfo {author} {\bibfnamefont {R.}~\bibnamefont {Nair}},\ and\ \bibinfo {author} {\bibfnamefont {X.-M.}\ \bibnamefont {Lu}},\ }\bibfield  {title} {\bibinfo {title} {Quantum theory of superresolution for two incoherent optical point sources},\ }\href {https://doi.org/10.1103/PhysRevX.6.031033} {\bibfield  {journal} {\bibinfo  {journal} {Physical Review X}\ }\textbf {\bibinfo {volume} {6}},\ \bibinfo {pages} {031033} (\bibinfo {year} {2016})}\BibitemShut {NoStop}%
\bibitem [{\citenamefont {Paúr}\ \emph {et~al.}(2016)\citenamefont {Paúr}, \citenamefont {Stoklasa}, \citenamefont {Hradil}, \citenamefont {Sánchez-Soto},\ and\ \citenamefont {Rehacek}}]{Pa_r_2016}%
  \BibitemOpen
  \bibfield  {author} {\bibinfo {author} {\bibfnamefont {M.}~\bibnamefont {Paúr}}, \bibinfo {author} {\bibfnamefont {B.}~\bibnamefont {Stoklasa}}, \bibinfo {author} {\bibfnamefont {Z.}~\bibnamefont {Hradil}}, \bibinfo {author} {\bibfnamefont {L.~L.}\ \bibnamefont {Sánchez-Soto}},\ and\ \bibinfo {author} {\bibfnamefont {J.}~\bibnamefont {Rehacek}},\ }\bibfield  {title} {\bibinfo {title} {Achieving the ultimate optical resolution},\ }\href {https://doi.org/10.1364/OPTICA.3.001144} {\bibfield  {journal} {\bibinfo  {journal} {Optica}\ }\textbf {\bibinfo {volume} {3}},\ \bibinfo {pages} {1144} (\bibinfo {year} {2016})}\BibitemShut {NoStop}%
\bibitem [{\citenamefont {Rehacek}\ \emph {et~al.}(2017)\citenamefont {Rehacek}, \citenamefont {Paúr}, \citenamefont {Stoklasa}, \citenamefont {Hradil},\ and\ \citenamefont {Sánchez-Soto}}]{Rehacek_2017}%
  \BibitemOpen
  \bibfield  {author} {\bibinfo {author} {\bibfnamefont {J.}~\bibnamefont {Rehacek}}, \bibinfo {author} {\bibfnamefont {M.}~\bibnamefont {Paúr}}, \bibinfo {author} {\bibfnamefont {B.}~\bibnamefont {Stoklasa}}, \bibinfo {author} {\bibfnamefont {Z.}~\bibnamefont {Hradil}},\ and\ \bibinfo {author} {\bibfnamefont {L.~L.}\ \bibnamefont {Sánchez-Soto}},\ }\bibfield  {title} {\bibinfo {title} {Optimal measurements for resolution beyond the rayleigh limit},\ }\href {https://doi.org/10.1364/OL.42.000231} {\bibfield  {journal} {\bibinfo  {journal} {Optics Letters}\ }\textbf {\bibinfo {volume} {42}},\ \bibinfo {pages} {231} (\bibinfo {year} {2017})}\BibitemShut {NoStop}%
\bibitem [{\citenamefont {Tham}\ \emph {et~al.}(2017)\citenamefont {Tham}, \citenamefont {Ferretti},\ and\ \citenamefont {Steinberg}}]{Tham_2017}%
  \BibitemOpen
  \bibfield  {author} {\bibinfo {author} {\bibfnamefont {W.-K.}\ \bibnamefont {Tham}}, \bibinfo {author} {\bibfnamefont {H.}~\bibnamefont {Ferretti}},\ and\ \bibinfo {author} {\bibfnamefont {A.~M.}\ \bibnamefont {Steinberg}},\ }\bibfield  {title} {\bibinfo {title} {Beating rayleigh’s curse by imaging using phase information},\ }\href {https://doi.org/10.1103/PhysRevLett.118.070801} {\bibfield  {journal} {\bibinfo  {journal} {Physical Review Letters}\ }\textbf {\bibinfo {volume} {118}},\ \bibinfo {pages} {070801} (\bibinfo {year} {2017})}\BibitemShut {NoStop}%
\bibitem [{\citenamefont {Gefen}\ \emph {et~al.}(2019)\citenamefont {Gefen}, \citenamefont {Rotem},\ and\ \citenamefont {Retzker}}]{Gefen_2019}%
  \BibitemOpen
  \bibfield  {author} {\bibinfo {author} {\bibfnamefont {T.}~\bibnamefont {Gefen}}, \bibinfo {author} {\bibfnamefont {A.}~\bibnamefont {Rotem}},\ and\ \bibinfo {author} {\bibfnamefont {A.}~\bibnamefont {Retzker}},\ }\bibfield  {title} {\bibinfo {title} {Overcoming resolution limits with quantum sensing},\ }\href {https://doi.org/10.1038/s41467-019-12817-y} {\bibfield  {journal} {\bibinfo  {journal} {Nature Communications}\ }\textbf {\bibinfo {volume} {10}},\ \bibinfo {pages} {4992} (\bibinfo {year} {2019})}\BibitemShut {NoStop}%
\bibitem [{\citenamefont {Pushkina}\ \emph {et~al.}(2021)\citenamefont {Pushkina}, \citenamefont {Maltese}, \citenamefont {Costa-Filho}, \citenamefont {Patel},\ and\ \citenamefont {Lvovsky}}]{Pushkina_2021}%
  \BibitemOpen
  \bibfield  {author} {\bibinfo {author} {\bibfnamefont {A.}~\bibnamefont {Pushkina}}, \bibinfo {author} {\bibfnamefont {G.}~\bibnamefont {Maltese}}, \bibinfo {author} {\bibfnamefont {J.}~\bibnamefont {Costa-Filho}}, \bibinfo {author} {\bibfnamefont {P.}~\bibnamefont {Patel}},\ and\ \bibinfo {author} {\bibfnamefont {A.}~\bibnamefont {Lvovsky}},\ }\bibfield  {title} {\bibinfo {title} {Superresolution linear optical imaging in the far field},\ }\href {https://doi.org/10.1103/PhysRevLett.127.253602} {\bibfield  {journal} {\bibinfo  {journal} {Physical Review Letters}\ }\textbf {\bibinfo {volume} {127}},\ \bibinfo {pages} {253602} (\bibinfo {year} {2021})}\BibitemShut {NoStop}%
\bibitem [{\citenamefont {Zanforlin}\ \emph {et~al.}(2022)\citenamefont {Zanforlin}, \citenamefont {Lupo}, \citenamefont {Connolly}, \citenamefont {Kok}, \citenamefont {Buller},\ and\ \citenamefont {Huang}}]{Zanforlin_2022}%
  \BibitemOpen
  \bibfield  {author} {\bibinfo {author} {\bibfnamefont {U.}~\bibnamefont {Zanforlin}}, \bibinfo {author} {\bibfnamefont {C.}~\bibnamefont {Lupo}}, \bibinfo {author} {\bibfnamefont {P.~W.~R.}\ \bibnamefont {Connolly}}, \bibinfo {author} {\bibfnamefont {P.}~\bibnamefont {Kok}}, \bibinfo {author} {\bibfnamefont {G.~S.}\ \bibnamefont {Buller}},\ and\ \bibinfo {author} {\bibfnamefont {Z.}~\bibnamefont {Huang}},\ }\bibfield  {title} {\bibinfo {title} {Optical quantum super-resolution imaging and hypothesis testing},\ }\href {https://doi.org/10.1038/s41467-022-32977-8} {\bibfield  {journal} {\bibinfo  {journal} {Nature Communications}\ }\textbf {\bibinfo {volume} {13}},\ \bibinfo {pages} {5373} (\bibinfo {year} {2022})}\BibitemShut {NoStop}%
\bibitem [{\citenamefont {Bonsma-Fisher}\ \emph {et~al.}(2019)\citenamefont {Bonsma-Fisher}, \citenamefont {Tham}, \citenamefont {Ferretti},\ and\ \citenamefont {Steinberg}}]{Bonsma_Fisher_2019}%
  \BibitemOpen
  \bibfield  {author} {\bibinfo {author} {\bibfnamefont {K.~A.~G.}\ \bibnamefont {Bonsma-Fisher}}, \bibinfo {author} {\bibfnamefont {W.-K.}\ \bibnamefont {Tham}}, \bibinfo {author} {\bibfnamefont {H.}~\bibnamefont {Ferretti}},\ and\ \bibinfo {author} {\bibfnamefont {A.~M.}\ \bibnamefont {Steinberg}},\ }\bibfield  {title} {\bibinfo {title} {Realistic sub-rayleigh imaging with phase-sensitive measurements},\ }\href {https://doi.org/10.1088/1367-2630/ab3d97} {\bibfield  {journal} {\bibinfo  {journal} {New Journal of Physics}\ }\textbf {\bibinfo {volume} {21}},\ \bibinfo {pages} {093010} (\bibinfo {year} {2019})}\BibitemShut {NoStop}%
\bibitem [{\citenamefont {Gessner}\ \emph {et~al.}(2020)\citenamefont {Gessner}, \citenamefont {Fabre},\ and\ \citenamefont {Treps}}]{Gessner_2020}%
  \BibitemOpen
  \bibfield  {author} {\bibinfo {author} {\bibfnamefont {M.}~\bibnamefont {Gessner}}, \bibinfo {author} {\bibfnamefont {C.}~\bibnamefont {Fabre}},\ and\ \bibinfo {author} {\bibfnamefont {N.}~\bibnamefont {Treps}},\ }\bibfield  {title} {\bibinfo {title} {Superresolution limits from measurement crosstalk},\ }\href {https://doi.org/10.1103/PhysRevLett.125.100501} {\bibfield  {journal} {\bibinfo  {journal} {Physical Review Letters}\ }\textbf {\bibinfo {volume} {125}},\ \bibinfo {pages} {100501} (\bibinfo {year} {2020})}\BibitemShut {NoStop}%
\bibitem [{\citenamefont {Lupo}(2020)}]{Lupo_2020}%
  \BibitemOpen
  \bibfield  {author} {\bibinfo {author} {\bibfnamefont {C.}~\bibnamefont {Lupo}},\ }\bibfield  {title} {\bibinfo {title} {Subwavelength quantum imaging with noisy detectors},\ }\href {https://doi.org/10.1103/PhysRevA.101.022323} {\bibfield  {journal} {\bibinfo  {journal} {Physical Review A}\ }\textbf {\bibinfo {volume} {101}},\ \bibinfo {pages} {022323} (\bibinfo {year} {2020})}\BibitemShut {NoStop}%
\bibitem [{\citenamefont {Sorelli}\ \emph {et~al.}(2021)\citenamefont {Sorelli}, \citenamefont {Gessner}, \citenamefont {Walschaers},\ and\ \citenamefont {Treps}}]{Sorelli_2021}%
  \BibitemOpen
  \bibfield  {author} {\bibinfo {author} {\bibfnamefont {G.}~\bibnamefont {Sorelli}}, \bibinfo {author} {\bibfnamefont {M.}~\bibnamefont {Gessner}}, \bibinfo {author} {\bibfnamefont {M.}~\bibnamefont {Walschaers}},\ and\ \bibinfo {author} {\bibfnamefont {N.}~\bibnamefont {Treps}},\ }\bibfield  {title} {\bibinfo {title} {Optimal observables and estimators for practical superresolution imaging},\ }\href {https://doi.org/10.1103/PhysRevLett.127.123604} {\bibfield  {journal} {\bibinfo  {journal} {Physical Review Letters}\ }\textbf {\bibinfo {volume} {127}},\ \bibinfo {pages} {123604} (\bibinfo {year} {2021})}\BibitemShut {NoStop}%
\bibitem [{\citenamefont {Oh}\ \emph {et~al.}(2021)\citenamefont {Oh}, \citenamefont {Zhou}, \citenamefont {Wong},\ and\ \citenamefont {Jiang}}]{Oh_2021}%
  \BibitemOpen
  \bibfield  {author} {\bibinfo {author} {\bibfnamefont {C.}~\bibnamefont {Oh}}, \bibinfo {author} {\bibfnamefont {S.}~\bibnamefont {Zhou}}, \bibinfo {author} {\bibfnamefont {Y.}~\bibnamefont {Wong}},\ and\ \bibinfo {author} {\bibfnamefont {L.}~\bibnamefont {Jiang}},\ }\bibfield  {title} {\bibinfo {title} {Quantum limits of superresolution in a noisy environment},\ }\href {https://doi.org/10.1103/PhysRevLett.126.120502} {\bibfield  {journal} {\bibinfo  {journal} {Physical Review Letters}\ }\textbf {\bibinfo {volume} {126}},\ \bibinfo {pages} {120502} (\bibinfo {year} {2021})}\BibitemShut {NoStop}%
\bibitem [{\citenamefont {Rouvière}\ \emph {et~al.}(2024)\citenamefont {Rouvière}, \citenamefont {Barral}, \citenamefont {Grateau}, \citenamefont {Karuseichyk}, \citenamefont {Sorelli}, \citenamefont {Walschaers},\ and\ \citenamefont {Treps}}]{Rouvi_re_2024}%
  \BibitemOpen
  \bibfield  {author} {\bibinfo {author} {\bibfnamefont {C.}~\bibnamefont {Rouvière}}, \bibinfo {author} {\bibfnamefont {D.}~\bibnamefont {Barral}}, \bibinfo {author} {\bibfnamefont {A.}~\bibnamefont {Grateau}}, \bibinfo {author} {\bibfnamefont {I.}~\bibnamefont {Karuseichyk}}, \bibinfo {author} {\bibfnamefont {G.}~\bibnamefont {Sorelli}}, \bibinfo {author} {\bibfnamefont {M.}~\bibnamefont {Walschaers}},\ and\ \bibinfo {author} {\bibfnamefont {N.}~\bibnamefont {Treps}},\ }\bibfield  {title} {\bibinfo {title} {Ultra-sensitive separation estimation of optical sources},\ }\href {https://doi.org/10.1364/OPTICA.500039} {\bibfield  {journal} {\bibinfo  {journal} {Optica}\ }\textbf {\bibinfo {volume} {11}},\ \bibinfo {pages} {166} (\bibinfo {year} {2024})}\BibitemShut {NoStop}%
\bibitem [{\citenamefont {Raymer}\ and\ \citenamefont {Banaszek}(2020)}]{Raymer_2020}%
  \BibitemOpen
  \bibfield  {author} {\bibinfo {author} {\bibfnamefont {M.~G.}\ \bibnamefont {Raymer}}\ and\ \bibinfo {author} {\bibfnamefont {K.}~\bibnamefont {Banaszek}},\ }\bibfield  {title} {\bibinfo {title} {Time-frequency optical filtering: efficiency vs. temporal-mode discrimination in incoherent and coherent implementations},\ }\href {https://doi.org/10.1364/OE.405618} {\bibfield  {journal} {\bibinfo  {journal} {Optics Express}\ }\textbf {\bibinfo {volume} {28}},\ \bibinfo {pages} {32819} (\bibinfo {year} {2020})}\BibitemShut {NoStop}%
\bibitem [{\citenamefont {Humphreys}\ \emph {et~al.}(2014)\citenamefont {Humphreys}, \citenamefont {Kolthammer}, \citenamefont {Nunn}, \citenamefont {Barbieri}, \citenamefont {Datta},\ and\ \citenamefont {Walmsley}}]{Humphreys_2014}%
  \BibitemOpen
  \bibfield  {author} {\bibinfo {author} {\bibfnamefont {P.~C.}\ \bibnamefont {Humphreys}}, \bibinfo {author} {\bibfnamefont {W.~S.}\ \bibnamefont {Kolthammer}}, \bibinfo {author} {\bibfnamefont {J.}~\bibnamefont {Nunn}}, \bibinfo {author} {\bibfnamefont {M.}~\bibnamefont {Barbieri}}, \bibinfo {author} {\bibfnamefont {A.}~\bibnamefont {Datta}},\ and\ \bibinfo {author} {\bibfnamefont {I.~A.}\ \bibnamefont {Walmsley}},\ }\bibfield  {title} {\bibinfo {title} {Continuous-variable quantum computing in optical time-frequency modes using quantum memories},\ }\href {https://doi.org/10.1103/PhysRevLett.113.130502} {\bibfield  {journal} {\bibinfo  {journal} {Physical Review Letters}\ }\textbf {\bibinfo {volume} {113}},\ \bibinfo {pages} {130502} (\bibinfo {year} {2014})}\BibitemShut {NoStop}%
\bibitem [{\citenamefont {Awschalom}\ \emph {et~al.}(2021)\citenamefont {Awschalom}, \citenamefont {Berggren}, \citenamefont {Bernien}, \citenamefont {Bhave}, \citenamefont {Carr}, \citenamefont {Davids}, \citenamefont {Economou}, \citenamefont {Englund}, \citenamefont {Faraon}, \citenamefont {Fejer}, \citenamefont {Guha}, \citenamefont {Gustafsson}, \citenamefont {Hu}, \citenamefont {Jiang}, \citenamefont {Kim}, \citenamefont {Korzh}, \citenamefont {Kumar}, \citenamefont {Kwiat}, \citenamefont {Lončar}, \citenamefont {Lukin}, \citenamefont {Miller}, \citenamefont {Monroe}, \citenamefont {Nam}, \citenamefont {Narang}, \citenamefont {Orcutt}, \citenamefont {Raymer}, \citenamefont {Safavi-Naeini}, \citenamefont {Spiropulu}, \citenamefont {Srinivasan}, \citenamefont {Sun}, \citenamefont {Vučković}, \citenamefont {Waks}, \citenamefont {Walsworth}, \citenamefont {Weiner},\ and\ \citenamefont {Zhang}}]{Awschalom_2021}%
  \BibitemOpen
  \bibfield  {author} {\bibinfo {author} {\bibfnamefont {D.}~\bibnamefont {Awschalom}}, \bibinfo {author} {\bibfnamefont {K.~K.}\ \bibnamefont {Berggren}}, \bibinfo {author} {\bibfnamefont {H.}~\bibnamefont {Bernien}}, \bibinfo {author} {\bibfnamefont {S.}~\bibnamefont {Bhave}}, \bibinfo {author} {\bibfnamefont {L.~D.}\ \bibnamefont {Carr}}, \bibinfo {author} {\bibfnamefont {P.}~\bibnamefont {Davids}}, \bibinfo {author} {\bibfnamefont {S.~E.}\ \bibnamefont {Economou}}, \bibinfo {author} {\bibfnamefont {D.}~\bibnamefont {Englund}}, \bibinfo {author} {\bibfnamefont {A.}~\bibnamefont {Faraon}}, \bibinfo {author} {\bibfnamefont {M.}~\bibnamefont {Fejer}}, \bibinfo {author} {\bibfnamefont {S.}~\bibnamefont {Guha}}, \bibinfo {author} {\bibfnamefont {M.~V.}\ \bibnamefont {Gustafsson}}, \bibinfo {author} {\bibfnamefont {E.}~\bibnamefont {Hu}}, \bibinfo {author} {\bibfnamefont {L.}~\bibnamefont {Jiang}}, \bibinfo {author} {\bibfnamefont {J.}~\bibnamefont {Kim}}, \bibinfo {author} {\bibfnamefont {B.}~\bibnamefont
  {Korzh}}, \bibinfo {author} {\bibfnamefont {P.}~\bibnamefont {Kumar}}, \bibinfo {author} {\bibfnamefont {P.~G.}\ \bibnamefont {Kwiat}}, \bibinfo {author} {\bibfnamefont {M.}~\bibnamefont {Lončar}}, \bibinfo {author} {\bibfnamefont {M.~D.}\ \bibnamefont {Lukin}}, \bibinfo {author} {\bibfnamefont {D.~A.}\ \bibnamefont {Miller}}, \bibinfo {author} {\bibfnamefont {C.}~\bibnamefont {Monroe}}, \bibinfo {author} {\bibfnamefont {S.~W.}\ \bibnamefont {Nam}}, \bibinfo {author} {\bibfnamefont {P.}~\bibnamefont {Narang}}, \bibinfo {author} {\bibfnamefont {J.~S.}\ \bibnamefont {Orcutt}}, \bibinfo {author} {\bibfnamefont {M.~G.}\ \bibnamefont {Raymer}}, \bibinfo {author} {\bibfnamefont {A.~H.}\ \bibnamefont {Safavi-Naeini}}, \bibinfo {author} {\bibfnamefont {M.}~\bibnamefont {Spiropulu}}, \bibinfo {author} {\bibfnamefont {K.}~\bibnamefont {Srinivasan}}, \bibinfo {author} {\bibfnamefont {S.}~\bibnamefont {Sun}}, \bibinfo {author} {\bibfnamefont {J.}~\bibnamefont {Vučković}}, \bibinfo {author} {\bibfnamefont
  {E.}~\bibnamefont {Waks}}, \bibinfo {author} {\bibfnamefont {R.}~\bibnamefont {Walsworth}}, \bibinfo {author} {\bibfnamefont {A.~M.}\ \bibnamefont {Weiner}},\ and\ \bibinfo {author} {\bibfnamefont {Z.}~\bibnamefont {Zhang}},\ }\bibfield  {title} {\bibinfo {title} {Development of quantum interconnects (quics) for next-generation information technologies},\ }\href {https://doi.org/10.1103/PRXQuantum.2.017002} {\bibfield  {journal} {\bibinfo  {journal} {PRX Quantum}\ }\textbf {\bibinfo {volume} {2}},\ \bibinfo {pages} {017002} (\bibinfo {year} {2021})}\BibitemShut {NoStop}%
\bibitem [{\citenamefont {Makino}\ \emph {et~al.}(2016)\citenamefont {Makino}, \citenamefont {Hashimoto}, \citenamefont {Yoshikawa}, \citenamefont {Ohdan}, \citenamefont {Toyama}, \citenamefont {van Loock},\ and\ \citenamefont {Furusawa}}]{Makino_2016}%
  \BibitemOpen
  \bibfield  {author} {\bibinfo {author} {\bibfnamefont {K.}~\bibnamefont {Makino}}, \bibinfo {author} {\bibfnamefont {Y.}~\bibnamefont {Hashimoto}}, \bibinfo {author} {\bibfnamefont {J.-i.}\ \bibnamefont {Yoshikawa}}, \bibinfo {author} {\bibfnamefont {H.}~\bibnamefont {Ohdan}}, \bibinfo {author} {\bibfnamefont {T.}~\bibnamefont {Toyama}}, \bibinfo {author} {\bibfnamefont {P.}~\bibnamefont {van Loock}},\ and\ \bibinfo {author} {\bibfnamefont {A.}~\bibnamefont {Furusawa}},\ }\bibfield  {title} {\bibinfo {title} {Synchronization of optical photons for quantum information processing},\ }\href {https://doi.org/10.1126/sciadv.1501772} {\bibfield  {journal} {\bibinfo  {journal} {Science Advances}\ }\textbf {\bibinfo {volume} {2}},\ \bibinfo {pages} {e1501772} (\bibinfo {year} {2016})}\BibitemShut {NoStop}%
\bibitem [{\citenamefont {Davidson}\ \emph {et~al.}(2023{\natexlab{a}})\citenamefont {Davidson}, \citenamefont {Yogev}, \citenamefont {Poem},\ and\ \citenamefont {Firstenberg}}]{Davidson_2023}%
  \BibitemOpen
  \bibfield  {author} {\bibinfo {author} {\bibfnamefont {O.}~\bibnamefont {Davidson}}, \bibinfo {author} {\bibfnamefont {O.}~\bibnamefont {Yogev}}, \bibinfo {author} {\bibfnamefont {E.}~\bibnamefont {Poem}},\ and\ \bibinfo {author} {\bibfnamefont {O.}~\bibnamefont {Firstenberg}},\ }\bibfield  {title} {\bibinfo {title} {Single-photon synchronization with a room-temperature atomic quantum memory},\ }\href {https://doi.org/10.1103/PhysRevLett.131.033601} {\bibfield  {journal} {\bibinfo  {journal} {Physical Review Letters}\ }\textbf {\bibinfo {volume} {131}},\ \bibinfo {pages} {033601} (\bibinfo {year} {2023}{\natexlab{a}})}\BibitemShut {NoStop}%
\bibitem [{\citenamefont {Kielpinski}\ \emph {et~al.}(2011)\citenamefont {Kielpinski}, \citenamefont {Corney},\ and\ \citenamefont {Wiseman}}]{Kielpinski_2011}%
  \BibitemOpen
  \bibfield  {author} {\bibinfo {author} {\bibfnamefont {D.}~\bibnamefont {Kielpinski}}, \bibinfo {author} {\bibfnamefont {J.~F.}\ \bibnamefont {Corney}},\ and\ \bibinfo {author} {\bibfnamefont {H.~M.}\ \bibnamefont {Wiseman}},\ }\bibfield  {title} {\bibinfo {title} {Quantum optical waveform conversion},\ }\href {https://doi.org/10.1103/physrevlett.106.130501} {\bibfield  {journal} {\bibinfo  {journal} {Physical Review Letters}\ }\textbf {\bibinfo {volume} {106}},\ \bibinfo {pages} {130501} (\bibinfo {year} {2011})}\BibitemShut {NoStop}%
\bibitem [{\citenamefont {Radnaev}\ \emph {et~al.}(2010)\citenamefont {Radnaev}, \citenamefont {Dudin}, \citenamefont {Zhao}, \citenamefont {Jen}, \citenamefont {Jenkins}, \citenamefont {Kuzmich},\ and\ \citenamefont {Kennedy}}]{Radnaev_2010}%
  \BibitemOpen
  \bibfield  {author} {\bibinfo {author} {\bibfnamefont {A.~G.}\ \bibnamefont {Radnaev}}, \bibinfo {author} {\bibfnamefont {Y.~O.}\ \bibnamefont {Dudin}}, \bibinfo {author} {\bibfnamefont {R.}~\bibnamefont {Zhao}}, \bibinfo {author} {\bibfnamefont {H.~H.}\ \bibnamefont {Jen}}, \bibinfo {author} {\bibfnamefont {S.~D.}\ \bibnamefont {Jenkins}}, \bibinfo {author} {\bibfnamefont {A.}~\bibnamefont {Kuzmich}},\ and\ \bibinfo {author} {\bibfnamefont {T.~A.~B.}\ \bibnamefont {Kennedy}},\ }\bibfield  {title} {\bibinfo {title} {A quantum memory with telecom-wavelength conversion},\ }\href {https://doi.org/10.1038/nphys1773} {\bibfield  {journal} {\bibinfo  {journal} {Nature Physics}\ }\textbf {\bibinfo {volume} {6}},\ \bibinfo {pages} {894} (\bibinfo {year} {2010})}\BibitemShut {NoStop}%
\bibitem [{\citenamefont {Karpiński}\ \emph {et~al.}(2016)\citenamefont {Karpiński}, \citenamefont {Jachura}, \citenamefont {Wright},\ and\ \citenamefont {Smith}}]{Karpi_ski_2016}%
  \BibitemOpen
  \bibfield  {author} {\bibinfo {author} {\bibfnamefont {M.}~\bibnamefont {Karpiński}}, \bibinfo {author} {\bibfnamefont {M.}~\bibnamefont {Jachura}}, \bibinfo {author} {\bibfnamefont {L.~J.}\ \bibnamefont {Wright}},\ and\ \bibinfo {author} {\bibfnamefont {B.~J.}\ \bibnamefont {Smith}},\ }\bibfield  {title} {\bibinfo {title} {Bandwidth manipulation of quantum light by an electro-optic time lens},\ }\href {https://doi.org/10.1038/nphoton.2016.228} {\bibfield  {journal} {\bibinfo  {journal} {Nature Photonics}\ }\textbf {\bibinfo {volume} {11}},\ \bibinfo {pages} {53} (\bibinfo {year} {2016})}\BibitemShut {NoStop}%
\bibitem [{\citenamefont {Zhang}\ and\ \citenamefont {Zhuang}(2021)}]{Zhang_2021}%
  \BibitemOpen
  \bibfield  {author} {\bibinfo {author} {\bibfnamefont {Z.}~\bibnamefont {Zhang}}\ and\ \bibinfo {author} {\bibfnamefont {Q.}~\bibnamefont {Zhuang}},\ }\bibfield  {title} {\bibinfo {title} {Distributed quantum sensing},\ }\href {https://doi.org/10.1088/2058-9565/abd4c3} {\bibfield  {journal} {\bibinfo  {journal} {Quantum Science and Technology}\ }\textbf {\bibinfo {volume} {6}},\ \bibinfo {pages} {043001} (\bibinfo {year} {2021})}\BibitemShut {NoStop}%
\bibitem [{\citenamefont {Donohue}\ \emph {et~al.}(2018)\citenamefont {Donohue}, \citenamefont {Ansari}, \citenamefont {Řeháček}, \citenamefont {Hradil}, \citenamefont {Stoklasa}, \citenamefont {Paúr}, \citenamefont {Sánchez-Soto},\ and\ \citenamefont {Silberhorn}}]{Donohue_2018}%
  \BibitemOpen
  \bibfield  {author} {\bibinfo {author} {\bibfnamefont {J.}~\bibnamefont {Donohue}}, \bibinfo {author} {\bibfnamefont {V.}~\bibnamefont {Ansari}}, \bibinfo {author} {\bibfnamefont {J.}~\bibnamefont {Řeháček}}, \bibinfo {author} {\bibfnamefont {Z.}~\bibnamefont {Hradil}}, \bibinfo {author} {\bibfnamefont {B.}~\bibnamefont {Stoklasa}}, \bibinfo {author} {\bibfnamefont {M.}~\bibnamefont {Paúr}}, \bibinfo {author} {\bibfnamefont {L.}~\bibnamefont {Sánchez-Soto}},\ and\ \bibinfo {author} {\bibfnamefont {C.}~\bibnamefont {Silberhorn}},\ }\bibfield  {title} {\bibinfo {title} {Quantum-limited time-frequency estimation through mode-selective photon measurement},\ }\href {https://doi.org/10.1103/PhysRevLett.121.090501} {\bibfield  {journal} {\bibinfo  {journal} {Physical Review Letters}\ }\textbf {\bibinfo {volume} {121}},\ \bibinfo {pages} {090501} (\bibinfo {year} {2018})}\BibitemShut {NoStop}%
\bibitem [{\citenamefont {Ansari}\ \emph {et~al.}(2021)\citenamefont {Ansari}, \citenamefont {Brecht}, \citenamefont {Gil-Lopez}, \citenamefont {Donohue}, \citenamefont {Řeháček}, \citenamefont {Hradil}, \citenamefont {Sánchez-Soto},\ and\ \citenamefont {Silberhorn}}]{Ansari_2021}%
  \BibitemOpen
  \bibfield  {author} {\bibinfo {author} {\bibfnamefont {V.}~\bibnamefont {Ansari}}, \bibinfo {author} {\bibfnamefont {B.}~\bibnamefont {Brecht}}, \bibinfo {author} {\bibfnamefont {J.}~\bibnamefont {Gil-Lopez}}, \bibinfo {author} {\bibfnamefont {J.~M.}\ \bibnamefont {Donohue}}, \bibinfo {author} {\bibfnamefont {J.}~\bibnamefont {Řeháček}}, \bibinfo {author} {\bibfnamefont {Z.}~\bibnamefont {Hradil}}, \bibinfo {author} {\bibfnamefont {L.~L.}\ \bibnamefont {Sánchez-Soto}},\ and\ \bibinfo {author} {\bibfnamefont {C.}~\bibnamefont {Silberhorn}},\ }\bibfield  {title} {\bibinfo {title} {Achieving the ultimate quantum timing resolution},\ }\href {https://doi.org/10.1103/PRXQuantum.2.010301} {\bibfield  {journal} {\bibinfo  {journal} {PRX Quantum}\ }\textbf {\bibinfo {volume} {2}},\ \bibinfo {pages} {010301} (\bibinfo {year} {2021})}\BibitemShut {NoStop}%
\bibitem [{\citenamefont {Serino}\ \emph {et~al.}(2025)\citenamefont {Serino}, \citenamefont {Eigner}, \citenamefont {Brecht},\ and\ \citenamefont {Silberhorn}}]{Serino_2025}%
  \BibitemOpen
  \bibfield  {author} {\bibinfo {author} {\bibfnamefont {L.}~\bibnamefont {Serino}}, \bibinfo {author} {\bibfnamefont {C.}~\bibnamefont {Eigner}}, \bibinfo {author} {\bibfnamefont {B.}~\bibnamefont {Brecht}},\ and\ \bibinfo {author} {\bibfnamefont {C.}~\bibnamefont {Silberhorn}},\ }\bibfield  {title} {\bibinfo {title} {Programmable time-frequency mode-sorting of single photons with a multi-output quantum pulse gate},\ }\href {https://doi.org/10.1364/OE.544206} {\bibfield  {journal} {\bibinfo  {journal} {Optics Express}\ }\textbf {\bibinfo {volume} {33}},\ \bibinfo {pages} {5577} (\bibinfo {year} {2025})}\BibitemShut {NoStop}%
\bibitem [{\citenamefont {Mazelanik}\ \emph {et~al.}(2022)\citenamefont {Mazelanik}, \citenamefont {Leszczyński},\ and\ \citenamefont {Parniak}}]{Mazelanik_2022}%
  \BibitemOpen
  \bibfield  {author} {\bibinfo {author} {\bibfnamefont {M.}~\bibnamefont {Mazelanik}}, \bibinfo {author} {\bibfnamefont {A.}~\bibnamefont {Leszczyński}},\ and\ \bibinfo {author} {\bibfnamefont {M.}~\bibnamefont {Parniak}},\ }\bibfield  {title} {\bibinfo {title} {Optical-domain spectral super-resolution via a quantum-memory-based time-frequency processor},\ }\href {https://doi.org/10.1038/s41467-022-28066-5} {\bibfield  {journal} {\bibinfo  {journal} {Nature Communications}\ }\textbf {\bibinfo {volume} {13}},\ \bibinfo {pages} {691} (\bibinfo {year} {2022})}\BibitemShut {NoStop}%
\bibitem [{\citenamefont {Shah}\ and\ \citenamefont {Fan}(2021)}]{Shah_2021}%
  \BibitemOpen
  \bibfield  {author} {\bibinfo {author} {\bibfnamefont {M.}~\bibnamefont {Shah}}\ and\ \bibinfo {author} {\bibfnamefont {L.}~\bibnamefont {Fan}},\ }\bibfield  {title} {\bibinfo {title} {Frequency superresolution with spectrotemporal shaping of photons},\ }\href {https://doi.org/10.1103/PhysRevApplied.15.034071} {\bibfield  {journal} {\bibinfo  {journal} {Physical Review Applied}\ }\textbf {\bibinfo {volume} {15}},\ \bibinfo {pages} {034071} (\bibinfo {year} {2021})}\BibitemShut {NoStop}%
\bibitem [{\citenamefont {Lipka}\ and\ \citenamefont {Parniak}(2024)}]{Lipka_2024}%
  \BibitemOpen
  \bibfield  {author} {\bibinfo {author} {\bibfnamefont {M.}~\bibnamefont {Lipka}}\ and\ \bibinfo {author} {\bibfnamefont {M.}~\bibnamefont {Parniak}},\ }\bibfield  {title} {\bibinfo {title} {Super-resolution of ultrafast pulses via spectral inversion},\ }\href {https://doi.org/10.1364/OPTICA.522555} {\bibfield  {journal} {\bibinfo  {journal} {Optica}\ }\textbf {\bibinfo {volume} {11}},\ \bibinfo {pages} {1226} (\bibinfo {year} {2024})}\BibitemShut {NoStop}%
\bibitem [{\citenamefont {Lvovsky}\ \emph {et~al.}(2009)\citenamefont {Lvovsky}, \citenamefont {Sanders},\ and\ \citenamefont {Tittel}}]{Lvovsky_2009}%
  \BibitemOpen
  \bibfield  {author} {\bibinfo {author} {\bibfnamefont {A.~I.}\ \bibnamefont {Lvovsky}}, \bibinfo {author} {\bibfnamefont {B.~C.}\ \bibnamefont {Sanders}},\ and\ \bibinfo {author} {\bibfnamefont {W.}~\bibnamefont {Tittel}},\ }\bibfield  {title} {\bibinfo {title} {Optical quantum memory},\ }\href {https://doi.org/10.1038/nphoton.2009.231} {\bibfield  {journal} {\bibinfo  {journal} {Nature Photonics}\ }\textbf {\bibinfo {volume} {3}},\ \bibinfo {pages} {706} (\bibinfo {year} {2009})}\BibitemShut {NoStop}%
\bibitem [{\citenamefont {Hervas}\ \emph {et~al.}(2025)\citenamefont {Hervas}, \citenamefont {Goldberg}, \citenamefont {Sanz}, \citenamefont {Hradil}, \citenamefont {Řeháček},\ and\ \citenamefont {Sánchez-Soto}}]{Hervas_2025}%
  \BibitemOpen
  \bibfield  {author} {\bibinfo {author} {\bibfnamefont {J.}~\bibnamefont {Hervas}}, \bibinfo {author} {\bibfnamefont {A.}~\bibnamefont {Goldberg}}, \bibinfo {author} {\bibfnamefont {A.}~\bibnamefont {Sanz}}, \bibinfo {author} {\bibfnamefont {Z.}~\bibnamefont {Hradil}}, \bibinfo {author} {\bibfnamefont {J.}~\bibnamefont {Řeháček}},\ and\ \bibinfo {author} {\bibfnamefont {L.}~\bibnamefont {Sánchez-Soto}},\ }\bibfield  {title} {\bibinfo {title} {Beyond the quantum cramér-rao bound},\ }\href {https://doi.org/10.1103/PhysRevLett.134.010804} {\bibfield  {journal} {\bibinfo  {journal} {Physical Review Letters}\ }\textbf {\bibinfo {volume} {134}},\ \bibinfo {pages} {010804} (\bibinfo {year} {2025})}\BibitemShut {NoStop}%
\bibitem [{\citenamefont {Cover}\ and\ \citenamefont {Thomas}(2005)}]{Cover_2005}%
  \BibitemOpen
  \bibfield  {author} {\bibinfo {author} {\bibfnamefont {T.~M.}\ \bibnamefont {Cover}}\ and\ \bibinfo {author} {\bibfnamefont {J.~A.}\ \bibnamefont {Thomas}},\ }\href {https://doi.org/10.1002/047174882x} {\emph {\bibinfo {title} {Elements of Information Theory}}},\ \bibinfo {edition} {second edition}\ ed.\ (\bibinfo  {publisher} {Wiley},\ \bibinfo {year} {2005})\BibitemShut {NoStop}%
\bibitem [{\citenamefont {Nunn}\ \emph {et~al.}(2008)\citenamefont {Nunn}, \citenamefont {Reim}, \citenamefont {Lee}, \citenamefont {Lorenz}, \citenamefont {Sussman}, \citenamefont {Walmsley},\ and\ \citenamefont {Jaksch}}]{Nunn_2008}%
  \BibitemOpen
  \bibfield  {author} {\bibinfo {author} {\bibfnamefont {J.}~\bibnamefont {Nunn}}, \bibinfo {author} {\bibfnamefont {K.}~\bibnamefont {Reim}}, \bibinfo {author} {\bibfnamefont {K.~C.}\ \bibnamefont {Lee}}, \bibinfo {author} {\bibfnamefont {V.~O.}\ \bibnamefont {Lorenz}}, \bibinfo {author} {\bibfnamefont {B.~J.}\ \bibnamefont {Sussman}}, \bibinfo {author} {\bibfnamefont {I.~A.}\ \bibnamefont {Walmsley}},\ and\ \bibinfo {author} {\bibfnamefont {D.}~\bibnamefont {Jaksch}},\ }\bibfield  {title} {\bibinfo {title} {Multimode memories in atomic ensembles},\ }\href {https://doi.org/10.1103/PhysRevLett.101.260502} {\bibfield  {journal} {\bibinfo  {journal} {Phys. Rev. Lett.}\ }\textbf {\bibinfo {volume} {101}},\ \bibinfo {pages} {260502} (\bibinfo {year} {2008})}\BibitemShut {NoStop}%
\bibitem [{\citenamefont {Řehaček}\ \emph {et~al.}(2017)\citenamefont {Řehaček}, \citenamefont {Hradil}, \citenamefont {Stoklasa}, \citenamefont {Paúr}, \citenamefont {Grover}, \citenamefont {Krzic},\ and\ \citenamefont {Sánchez-Soto}}]{_eha_ek_2017}%
  \BibitemOpen
  \bibfield  {author} {\bibinfo {author} {\bibfnamefont {J.}~\bibnamefont {Řehaček}}, \bibinfo {author} {\bibfnamefont {Z.}~\bibnamefont {Hradil}}, \bibinfo {author} {\bibfnamefont {B.}~\bibnamefont {Stoklasa}}, \bibinfo {author} {\bibfnamefont {M.}~\bibnamefont {Paúr}}, \bibinfo {author} {\bibfnamefont {J.}~\bibnamefont {Grover}}, \bibinfo {author} {\bibfnamefont {A.}~\bibnamefont {Krzic}},\ and\ \bibinfo {author} {\bibfnamefont {L.~L.}\ \bibnamefont {Sánchez-Soto}},\ }\bibfield  {title} {\bibinfo {title} {Multiparameter quantum metrology of incoherent point sources: Towards realistic superresolution},\ }\href {https://doi.org/10.1103/PhysRevA.96.062107} {\bibfield  {journal} {\bibinfo  {journal} {Physical Review A}\ }\textbf {\bibinfo {volume} {96}},\ \bibinfo {pages} {062107} (\bibinfo {year} {2017})}\BibitemShut {NoStop}%
\bibitem [{\citenamefont {Liu}\ \emph {et~al.}(2019)\citenamefont {Liu}, \citenamefont {Yuan}, \citenamefont {Lu},\ and\ \citenamefont {Wang}}]{Liu_2019}%
  \BibitemOpen
  \bibfield  {author} {\bibinfo {author} {\bibfnamefont {J.}~\bibnamefont {Liu}}, \bibinfo {author} {\bibfnamefont {H.}~\bibnamefont {Yuan}}, \bibinfo {author} {\bibfnamefont {X.-M.}\ \bibnamefont {Lu}},\ and\ \bibinfo {author} {\bibfnamefont {X.}~\bibnamefont {Wang}},\ }\bibfield  {title} {\bibinfo {title} {Quantum fisher information matrix and multiparameter estimation},\ }\href {https://doi.org/10.1088/1751-8121/ab5d4d} {\bibfield  {journal} {\bibinfo  {journal} {Journal of Physics A: Mathematical and Theoretical}\ }\textbf {\bibinfo {volume} {53}},\ \bibinfo {pages} {023001} (\bibinfo {year} {2019})}\BibitemShut {NoStop}%
\bibitem [{\citenamefont {Shao}\ and\ \citenamefont {Lu}(2022)}]{Shao_2022}%
  \BibitemOpen
  \bibfield  {author} {\bibinfo {author} {\bibfnamefont {J.}~\bibnamefont {Shao}}\ and\ \bibinfo {author} {\bibfnamefont {X.-M.}\ \bibnamefont {Lu}},\ }\bibfield  {title} {\bibinfo {title} {Performance-tradeoff relation for locating two incoherent optical point sources},\ }\href {https://doi.org/10.1103/PhysRevA.105.062416} {\bibfield  {journal} {\bibinfo  {journal} {Physical Review A}\ }\textbf {\bibinfo {volume} {105}},\ \bibinfo {pages} {062416} (\bibinfo {year} {2022})}\BibitemShut {NoStop}%
\bibitem [{\citenamefont {Gosalia}\ and\ \citenamefont {Malaney}(2025)}]{Gosalia_2025}%
  \BibitemOpen
  \bibfield  {author} {\bibinfo {author} {\bibfnamefont {R.~K.}\ \bibnamefont {Gosalia}}\ and\ \bibinfo {author} {\bibfnamefont {R.}~\bibnamefont {Malaney}},\ }\bibfield  {title} {\bibinfo {title} {Quantum-enhanced clock synchronization using prior statistical information},\ }\href {https://doi.org/10.1088/2058-9565/adad92} {\bibfield  {journal} {\bibinfo  {journal} {Quantum Science and Technology}\ }\textbf {\bibinfo {volume} {10}},\ \bibinfo {pages} {025022} (\bibinfo {year} {2025})}\BibitemShut {NoStop}%
\bibitem [{\citenamefont {Giovannetti}\ \emph {et~al.}(2001)\citenamefont {Giovannetti}, \citenamefont {Lloyd},\ and\ \citenamefont {Maccone}}]{Giovannetti_2001}%
  \BibitemOpen
  \bibfield  {author} {\bibinfo {author} {\bibfnamefont {V.}~\bibnamefont {Giovannetti}}, \bibinfo {author} {\bibfnamefont {S.}~\bibnamefont {Lloyd}},\ and\ \bibinfo {author} {\bibfnamefont {L.}~\bibnamefont {Maccone}},\ }\bibfield  {title} {\bibinfo {title} {Quantum-enhanced positioning and clock synchronization},\ }\href {https://doi.org/10.1038/35086525} {\bibfield  {journal} {\bibinfo  {journal} {Nature}\ }\textbf {\bibinfo {volume} {412}},\ \bibinfo {pages} {417} (\bibinfo {year} {2001})}\BibitemShut {NoStop}%
\bibitem [{\citenamefont {Lamine}\ \emph {et~al.}(2008)\citenamefont {Lamine}, \citenamefont {Fabre},\ and\ \citenamefont {Treps}}]{Lamine_2008}%
  \BibitemOpen
  \bibfield  {author} {\bibinfo {author} {\bibfnamefont {B.}~\bibnamefont {Lamine}}, \bibinfo {author} {\bibfnamefont {C.}~\bibnamefont {Fabre}},\ and\ \bibinfo {author} {\bibfnamefont {N.}~\bibnamefont {Treps}},\ }\bibfield  {title} {\bibinfo {title} {Quantum improvement of time transfer between remote clocks},\ }\href {https://doi.org/10.1103/PhysRevLett.101.123601} {\bibfield  {journal} {\bibinfo  {journal} {Physical Review Letters}\ }\textbf {\bibinfo {volume} {101}},\ \bibinfo {pages} {123601} (\bibinfo {year} {2008})}\BibitemShut {NoStop}%
\bibitem [{\citenamefont {Kruse}\ \emph {et~al.}(2023)\citenamefont {Kruse}, \citenamefont {Serino}, \citenamefont {Folge}, \citenamefont {Oviedo}, \citenamefont {Bhattacharjee}, \citenamefont {Stefszky}, \citenamefont {Scheytt}, \citenamefont {Brecht},\ and\ \citenamefont {Silberhorn}}]{Kruse_2023}%
  \BibitemOpen
  \bibfield  {author} {\bibinfo {author} {\bibfnamefont {S.}~\bibnamefont {Kruse}}, \bibinfo {author} {\bibfnamefont {L.}~\bibnamefont {Serino}}, \bibinfo {author} {\bibfnamefont {P.}~\bibnamefont {Folge}}, \bibinfo {author} {\bibfnamefont {D.~E.}\ \bibnamefont {Oviedo}}, \bibinfo {author} {\bibfnamefont {A.}~\bibnamefont {Bhattacharjee}}, \bibinfo {author} {\bibfnamefont {M.}~\bibnamefont {Stefszky}}, \bibinfo {author} {\bibfnamefont {J.~C.}\ \bibnamefont {Scheytt}}, \bibinfo {author} {\bibfnamefont {B.}~\bibnamefont {Brecht}},\ and\ \bibinfo {author} {\bibfnamefont {C.}~\bibnamefont {Silberhorn}},\ }\bibfield  {title} {\bibinfo {title} {A pulsed lidar system with ultimate quantum range accuracy},\ }\href {https://doi.org/10.1109/LPT.2023.3277515} {\bibfield  {journal} {\bibinfo  {journal} {IEEE Photonics Technology Letters}\ }\textbf {\bibinfo {volume} {35}},\ \bibinfo {pages} {769} (\bibinfo {year} {2023})}\BibitemShut {NoStop}%
\bibitem [{\citenamefont {Huang}\ \emph {et~al.}(2021)\citenamefont {Huang}, \citenamefont {Lupo},\ and\ \citenamefont {Kok}}]{Huang_2021}%
  \BibitemOpen
  \bibfield  {author} {\bibinfo {author} {\bibfnamefont {Z.}~\bibnamefont {Huang}}, \bibinfo {author} {\bibfnamefont {C.}~\bibnamefont {Lupo}},\ and\ \bibinfo {author} {\bibfnamefont {P.}~\bibnamefont {Kok}},\ }\bibfield  {title} {\bibinfo {title} {Quantum-limited estimation of range and velocity},\ }\href {https://doi.org/10.1103/PRXQuantum.2.030303} {\bibfield  {journal} {\bibinfo  {journal} {PRX Quantum}\ }\textbf {\bibinfo {volume} {2}},\ \bibinfo {pages} {030303} (\bibinfo {year} {2021})}\BibitemShut {NoStop}%
\bibitem [{\citenamefont {Burdekin}\ \emph {et~al.}(2025)\citenamefont {Burdekin}, \citenamefont {Maillette~de Buy~Wenniger}, \citenamefont {Sagona-Stophel}, \citenamefont {Szuniewicz}, \citenamefont {Zhang}, \citenamefont {Thomas},\ and\ \citenamefont {Walmsley}}]{Burdekin_2025}%
  \BibitemOpen
  \bibfield  {author} {\bibinfo {author} {\bibfnamefont {P.~M.}\ \bibnamefont {Burdekin}}, \bibinfo {author} {\bibfnamefont {I.}~\bibnamefont {Maillette~de Buy~Wenniger}}, \bibinfo {author} {\bibfnamefont {S.}~\bibnamefont {Sagona-Stophel}}, \bibinfo {author} {\bibfnamefont {J.}~\bibnamefont {Szuniewicz}}, \bibinfo {author} {\bibfnamefont {A.}~\bibnamefont {Zhang}}, \bibinfo {author} {\bibfnamefont {S.~E.}\ \bibnamefont {Thomas}},\ and\ \bibinfo {author} {\bibfnamefont {I.~A.}\ \bibnamefont {Walmsley}},\ }\bibfield  {title} {\bibinfo {title} {Enhancing quantum memories with light-matter interference},\ }\href {https://doi.org/10.1364/OPTICA.564780} {\bibfield  {journal} {\bibinfo  {journal} {Optica}\ }\textbf {\bibinfo {volume} {12}},\ \bibinfo {pages} {1514} (\bibinfo {year} {2025})}\BibitemShut {NoStop}%
\bibitem [{\citenamefont {Saunders}\ \emph {et~al.}(2016)\citenamefont {Saunders}, \citenamefont {Munns}, \citenamefont {Champion}, \citenamefont {Qiu}, \citenamefont {Kaczmarek}, \citenamefont {Poem}, \citenamefont {Ledingham}, \citenamefont {Walmsley},\ and\ \citenamefont {Nunn}}]{Saunders_2016}%
  \BibitemOpen
  \bibfield  {author} {\bibinfo {author} {\bibfnamefont {D.~J.}\ \bibnamefont {Saunders}}, \bibinfo {author} {\bibfnamefont {J.~H.~D.}\ \bibnamefont {Munns}}, \bibinfo {author} {\bibfnamefont {T.~F.~M.}\ \bibnamefont {Champion}}, \bibinfo {author} {\bibfnamefont {C.}~\bibnamefont {Qiu}}, \bibinfo {author} {\bibfnamefont {K.~T.}\ \bibnamefont {Kaczmarek}}, \bibinfo {author} {\bibfnamefont {E.}~\bibnamefont {Poem}}, \bibinfo {author} {\bibfnamefont {P.~M.}\ \bibnamefont {Ledingham}}, \bibinfo {author} {\bibfnamefont {I.~A.}\ \bibnamefont {Walmsley}},\ and\ \bibinfo {author} {\bibfnamefont {J.}~\bibnamefont {Nunn}},\ }\bibfield  {title} {\bibinfo {title} {Cavity-enhanced room-temperature broadband raman memory},\ }\href {https://doi.org/10.1103/PhysRevLett.116.090501} {\bibfield  {journal} {\bibinfo  {journal} {Phys. Rev. Lett.}\ }\textbf {\bibinfo {volume} {116}},\ \bibinfo {pages} {090501} (\bibinfo {year} {2016})}\BibitemShut {NoStop}%
\bibitem [{\citenamefont {Gorshkov}\ \emph {et~al.}(2008)\citenamefont {Gorshkov}, \citenamefont {Calarco}, \citenamefont {Lukin},\ and\ \citenamefont {S\o{}rensen}}]{Gorshkov_2008}%
  \BibitemOpen
  \bibfield  {author} {\bibinfo {author} {\bibfnamefont {A.~V.}\ \bibnamefont {Gorshkov}}, \bibinfo {author} {\bibfnamefont {T.}~\bibnamefont {Calarco}}, \bibinfo {author} {\bibfnamefont {M.~D.}\ \bibnamefont {Lukin}},\ and\ \bibinfo {author} {\bibfnamefont {A.~S.}\ \bibnamefont {S\o{}rensen}},\ }\bibfield  {title} {\bibinfo {title} {Photon storage in $\ensuremath{\Lambda}$-type optically dense atomic media. iv. optimal control using gradient ascent},\ }\href {https://doi.org/10.1103/PhysRevA.77.043806} {\bibfield  {journal} {\bibinfo  {journal} {Phys. Rev. A}\ }\textbf {\bibinfo {volume} {77}},\ \bibinfo {pages} {043806} (\bibinfo {year} {2008})}\BibitemShut {NoStop}%
\bibitem [{\citenamefont {Guo}\ \emph {et~al.}(2019)\citenamefont {Guo}, \citenamefont {Feng}, \citenamefont {Yang}, \citenamefont {Yu}, \citenamefont {Chen}, \citenamefont {Yuan},\ and\ \citenamefont {Zhang}}]{Guo_2019}%
  \BibitemOpen
  \bibfield  {author} {\bibinfo {author} {\bibfnamefont {J.}~\bibnamefont {Guo}}, \bibinfo {author} {\bibfnamefont {X.}~\bibnamefont {Feng}}, \bibinfo {author} {\bibfnamefont {P.}~\bibnamefont {Yang}}, \bibinfo {author} {\bibfnamefont {Z.}~\bibnamefont {Yu}}, \bibinfo {author} {\bibfnamefont {L.~Q.}\ \bibnamefont {Chen}}, \bibinfo {author} {\bibfnamefont {C.-H.}\ \bibnamefont {Yuan}},\ and\ \bibinfo {author} {\bibfnamefont {W.}~\bibnamefont {Zhang}},\ }\bibfield  {title} {\bibinfo {title} {High-performance raman quantum memory with optimal control in room temperature atoms},\ }\href {https://doi.org/10.1038/s41467-018-08118-5} {\bibfield  {journal} {\bibinfo  {journal} {Nature Communications}\ }\textbf {\bibinfo {volume} {10}},\ \bibinfo {pages} {148} (\bibinfo {year} {2019})}\BibitemShut {NoStop}%
\bibitem [{\citenamefont {Davidson}\ \emph {et~al.}(2023{\natexlab{b}})\citenamefont {Davidson}, \citenamefont {Yogev}, \citenamefont {Poem},\ and\ \citenamefont {Firstenberg}}]{Davidson_2023_FLAME}%
  \BibitemOpen
  \bibfield  {author} {\bibinfo {author} {\bibfnamefont {O.}~\bibnamefont {Davidson}}, \bibinfo {author} {\bibfnamefont {O.}~\bibnamefont {Yogev}}, \bibinfo {author} {\bibfnamefont {E.}~\bibnamefont {Poem}},\ and\ \bibinfo {author} {\bibfnamefont {O.}~\bibnamefont {Firstenberg}},\ }\bibfield  {title} {\bibinfo {title} {Fast, noise-free atomic optical memory with 35-percent end-to-end efficiency},\ }\href {https://doi.org/10.1038/s42005-023-01247-4} {\bibfield  {journal} {\bibinfo  {journal} {Communications Physics}\ }\textbf {\bibinfo {volume} {6}},\ \bibinfo {pages} {131} (\bibinfo {year} {2023}{\natexlab{b}})}\BibitemShut {NoStop}%
\bibitem [{\citenamefont {Kaczmarek}\ \emph {et~al.}(2018)\citenamefont {Kaczmarek}, \citenamefont {Ledingham}, \citenamefont {Brecht}, \citenamefont {Thomas}, \citenamefont {Thekkadath}, \citenamefont {Lazo-Arjona}, \citenamefont {Munns}, \citenamefont {Poem}, \citenamefont {Feizpour}, \citenamefont {Saunders}, \citenamefont {Nunn},\ and\ \citenamefont {Walmsley}}]{Kaczmarek_2018}%
  \BibitemOpen
  \bibfield  {author} {\bibinfo {author} {\bibfnamefont {K.~T.}\ \bibnamefont {Kaczmarek}}, \bibinfo {author} {\bibfnamefont {P.~M.}\ \bibnamefont {Ledingham}}, \bibinfo {author} {\bibfnamefont {B.}~\bibnamefont {Brecht}}, \bibinfo {author} {\bibfnamefont {S.~E.}\ \bibnamefont {Thomas}}, \bibinfo {author} {\bibfnamefont {G.~S.}\ \bibnamefont {Thekkadath}}, \bibinfo {author} {\bibfnamefont {O.}~\bibnamefont {Lazo-Arjona}}, \bibinfo {author} {\bibfnamefont {J.~H.~D.}\ \bibnamefont {Munns}}, \bibinfo {author} {\bibfnamefont {E.}~\bibnamefont {Poem}}, \bibinfo {author} {\bibfnamefont {A.}~\bibnamefont {Feizpour}}, \bibinfo {author} {\bibfnamefont {D.~J.}\ \bibnamefont {Saunders}}, \bibinfo {author} {\bibfnamefont {J.}~\bibnamefont {Nunn}},\ and\ \bibinfo {author} {\bibfnamefont {I.~A.}\ \bibnamefont {Walmsley}},\ }\bibfield  {title} {\bibinfo {title} {High-speed noise-free optical quantum memory},\ }\href {https://doi.org/10.1103/PhysRevA.97.042316} {\bibfield  {journal} {\bibinfo  {journal} {Phys. Rev. A}\
  }\textbf {\bibinfo {volume} {97}},\ \bibinfo {pages} {042316} (\bibinfo {year} {2018})}\BibitemShut {NoStop}%
\bibitem [{\citenamefont {Thomas}\ \emph {et~al.}(2023)\citenamefont {Thomas}, \citenamefont {Sagona-Stophel}, \citenamefont {Schofield}, \citenamefont {Walmsley},\ and\ \citenamefont {Ledingham}}]{Thomas_2023}%
  \BibitemOpen
  \bibfield  {author} {\bibinfo {author} {\bibfnamefont {S.}~\bibnamefont {Thomas}}, \bibinfo {author} {\bibfnamefont {S.}~\bibnamefont {Sagona-Stophel}}, \bibinfo {author} {\bibfnamefont {Z.}~\bibnamefont {Schofield}}, \bibinfo {author} {\bibfnamefont {I.}~\bibnamefont {Walmsley}},\ and\ \bibinfo {author} {\bibfnamefont {P.}~\bibnamefont {Ledingham}},\ }\bibfield  {title} {\bibinfo {title} {Single-photon-compatible telecommunications-band quantum memory in a hot atomic gas},\ }\href {https://doi.org/10.1103/PhysRevApplied.19.L031005} {\bibfield  {journal} {\bibinfo  {journal} {Phys. Rev. Appl.}\ }\textbf {\bibinfo {volume} {19}},\ \bibinfo {pages} {L031005} (\bibinfo {year} {2023})}\BibitemShut {NoStop}%
\bibitem [{\citenamefont {Thomas}\ \emph {et~al.}(2019)\citenamefont {Thomas}, \citenamefont {Hird}, \citenamefont {Munns}, \citenamefont {Brecht}, \citenamefont {Saunders}, \citenamefont {Nunn}, \citenamefont {Walmsley},\ and\ \citenamefont {Ledingham}}]{Thomas_2019}%
  \BibitemOpen
  \bibfield  {author} {\bibinfo {author} {\bibfnamefont {S.~E.}\ \bibnamefont {Thomas}}, \bibinfo {author} {\bibfnamefont {T.~M.}\ \bibnamefont {Hird}}, \bibinfo {author} {\bibfnamefont {J.~H.~D.}\ \bibnamefont {Munns}}, \bibinfo {author} {\bibfnamefont {B.}~\bibnamefont {Brecht}}, \bibinfo {author} {\bibfnamefont {D.~J.}\ \bibnamefont {Saunders}}, \bibinfo {author} {\bibfnamefont {J.}~\bibnamefont {Nunn}}, \bibinfo {author} {\bibfnamefont {I.~A.}\ \bibnamefont {Walmsley}},\ and\ \bibinfo {author} {\bibfnamefont {P.~M.}\ \bibnamefont {Ledingham}},\ }\bibfield  {title} {\bibinfo {title} {Raman quantum memory with built-in suppression of four-wave-mixing noise},\ }\href {https://doi.org/10.1103/PhysRevA.100.033801} {\bibfield  {journal} {\bibinfo  {journal} {Phys. Rev. A}\ }\textbf {\bibinfo {volume} {100}},\ \bibinfo {pages} {033801} (\bibinfo {year} {2019})}\BibitemShut {NoStop}%
\bibitem [{\citenamefont {Klein}\ \emph {et~al.}(2009)\citenamefont {Klein}, \citenamefont {Hohensee}, \citenamefont {Nemiroski}, \citenamefont {Xiao}, \citenamefont {Phillips},\ and\ \citenamefont {Walsworth}}]{Klein_2009}%
  \BibitemOpen
  \bibfield  {author} {\bibinfo {author} {\bibfnamefont {M.}~\bibnamefont {Klein}}, \bibinfo {author} {\bibfnamefont {M.}~\bibnamefont {Hohensee}}, \bibinfo {author} {\bibfnamefont {A.}~\bibnamefont {Nemiroski}}, \bibinfo {author} {\bibfnamefont {Y.}~\bibnamefont {Xiao}}, \bibinfo {author} {\bibfnamefont {D.~F.}\ \bibnamefont {Phillips}},\ and\ \bibinfo {author} {\bibfnamefont {R.~L.}\ \bibnamefont {Walsworth}},\ }\bibfield  {title} {\bibinfo {title} {Slow light in narrow paraffin-coated vapor cells},\ }\href {https://doi.org/10.1063/1.3207825} {\bibfield  {journal} {\bibinfo  {journal} {Applied Physics Letters}\ }\textbf {\bibinfo {volume} {95}},\ \bibinfo {pages} {091102} (\bibinfo {year} {2009})}\BibitemShut {NoStop}%
\bibitem [{\citenamefont {Frehlich}(1997)}]{Frehlich_1997}%
  \BibitemOpen
  \bibfield  {author} {\bibinfo {author} {\bibfnamefont {R.}~\bibnamefont {Frehlich}},\ }\bibfield  {title} {\bibinfo {title} {Effects of wind turbulence on coherent doppler lidar performance},\ }\href {https://doi.org/10.1175/1520-0426(1997)014<0054:EOWTOC>2.0.CO;2} {\bibfield  {journal} {\bibinfo  {journal} {Journal of Atmospheric and Oceanic Technology}\ }\textbf {\bibinfo {volume} {14}},\ \bibinfo {pages} {54} (\bibinfo {year} {1997})}\BibitemShut {NoStop}%
\bibitem [{\citenamefont {McGill}\ \emph {et~al.}(1997)\citenamefont {McGill}, \citenamefont {Skinner},\ and\ \citenamefont {Irgang}}]{McGill_1997}%
  \BibitemOpen
  \bibfield  {author} {\bibinfo {author} {\bibfnamefont {M.~J.}\ \bibnamefont {McGill}}, \bibinfo {author} {\bibfnamefont {W.~R.}\ \bibnamefont {Skinner}},\ and\ \bibinfo {author} {\bibfnamefont {T.~D.}\ \bibnamefont {Irgang}},\ }\bibfield  {title} {\bibinfo {title} {Validation of wind profiles measured with incoherent doppler lidar},\ }\href {https://doi.org/10.1364/ao.36.001928} {\bibfield  {journal} {\bibinfo  {journal} {Applied Optics}\ }\textbf {\bibinfo {volume} {36}},\ \bibinfo {pages} {1928} (\bibinfo {year} {1997})}\BibitemShut {NoStop}%
\end{thebibliography}
%

\begin{widetext}
\newpage
\setcounter{equation}{0}
\setcounter{figure}{0}
\renewcommand{\thefigure}{S\arabic{figure}}
\renewcommand{\theequation}{S\arabic{equation}}

\section*{Supplementary Information: Super-resolving frequency measurement with mode-selective quantum memory}
\subsection{Fundamental bias at small separations}

\par
In the maximum likelihood estimation process, the non-negativity constraint of the parameter space $\Theta$ introduces a positive bias at very small frequency separations, where the distribution of an ideally unbiased estimator may extend into the negative region. Here, we present both an analytical analysis and numerical simulations to characterise this fundamental bias of a two-mode demultiplexing system with crosstalk matrix $\boldsymbol{M}$ (Eq.~\ref{eq:crosstalk_matrix} in the main text).

In the absence of crosstalk, the ideal detection probability for mode $n$ is
\begin{equation}
    P(n|\epsilon)=\frac{\epsilon^{2n}}{16^nn!}\exp\left(-\frac{\epsilon^2}{16}\right).
    \label{eq:hgprob}    
\end{equation}
And the Fisher information for this HG measurement basis can be calculated as
\begin{equation}
    \mathcal{F}_{\text{HG}}(\epsilon)=\sum_{n=0}^{\infty}\frac{1}{P(n|\epsilon)}\left(\frac{\partial P(n|\epsilon)}{\partial\epsilon}\right)^2.
    \label{eq:FI_manymode}
\end{equation}
Considering only the first two modes, the normalised probability of projecting onto the HG$_1$ mode after perturbation by the crosstalk matrix $\boldsymbol{M}$ is
\begin{equation}
    \tilde{P}(1|\epsilon) = \beta-\frac{16(\alpha+\beta-1)}{16+\epsilon^2},
    \label{eq:HG0_prob_crosstalk}    
\end{equation}
and the HG$_0$ projection probability is $\tilde{P}(0|\epsilon)=1-\tilde{P}(1|\epsilon)$. The exact expression for the Fisher information of the non-ideal two-mode projection method can be derived from Eq.~\ref{eq:FI_manymode} in the main text as follows:
\begin{equation}
    \mathcal{F}(\epsilon)=\frac{(\alpha+\beta-1)^2(\epsilon/4)^2}{4(1+(\epsilon/4)^2)^2(\alpha+(1-\beta)(\epsilon/4)^2)(1-\alpha+\beta(\epsilon/4)^2)}.
    \label{eq:non_ideal_two_mode_FI}  
\end{equation}
We assume the mode crosstalk between the two modes is low, i.e. $\alpha\approx1$ and $\beta\approx1$, and the separation is small. Then Eq.~\ref{eq:non_ideal_two_mode_FI} can be approximated as Eq.~\ref{eq:FI_2mode_crosstalk} in the main text. Using maximum likelihood estimation (Eq.~\ref{eq:MLE_estimator} in the main text), the likelihood function is given by
\begin{equation}
    \mathcal{L}=N_0\ln{\tilde{P}(0|\epsilon)}+N_1\ln{\tilde{P}(1|\epsilon)},
    \label{eq:MLE_likelihood}    
\end{equation}
where $N_0$ and $N_1$ represent the experimentally measured detection counts for HG$_0$ and HG$_1$ projections, respectively, with the total count given by $N=N_0+N_1$. The analytical solution of the MLE estimator can be obtained by solving the critical point equation $d\mathcal{L}/d\epsilon=0$, yielding
$\hat{\epsilon}=\sqrt{\frac{16(N_1 - (1 - \alpha)N)}{\beta N - N_1}}$. However, this solution is valid only when both the numerator and denominator are non-negative; thus, we require $ N_1 \geq (1 - \alpha)N$ (We neglect the condition $N_1<\beta N$, as the probability of $N_1>\beta N$ is negligible for small separations and low mode crosstalk). This condition corresponds to the range of the modal prediction discussed in the main paper. If a combination of $N_0$ and $N_1$ values falls outside this range, the likelihood function $\mathcal{L}$ achieves its maximum at the non-differentiable boundary point $\epsilon=0$. In such cases, the MLE defaults to $\hat{\epsilon}=0$. Therefore, the MLE estimators are
\begin{equation}
    \hat{\epsilon}_\text{MLE} =
        \begin{cases} 
        \sqrt{\frac{16(N_1 - (1 - \alpha)N)}{\beta N - N_1}}, & \text{if }  N_1 \geq (1 - \alpha)N, \\
        0, & \text{if }  N_1<(1 - \alpha)N.
        \end{cases}
    \label{eq:MLE_estimator_supp}    
\end{equation}
The maximum likelihood estimation method inherently produces only non-negative estimators based on the detected counts. When $N_0$ and $N_1$ result in a negative estimate, the estimator is set to zero. Consequently, this introduces a bias toward positive values, as all negative estimates are truncated to zero.

Since we employ photon count detection, the detection counts $N_0$ and $N_1$ follow Poisson distributions. Specifically, we model $N_1$ as $\text{Poisson}(\mu_1)$, where $\mu_1=\tilde{P}(1|\epsilon)N$. Substituting this distribution into the estimator formula in Eq.~\ref{eq:MLE_estimator_supp}, we derive the mean square error as
\begin{equation}
    \text{MSE}(\epsilon, N) = \sum_{k=0}^{(1-\alpha)N-1} \epsilon^2 \cdot \frac{\mu_1^k e^{-\mu_1}}{k!} +\sum_{k=(1-\alpha)N}^{\beta N-1} \left( 4 \sqrt{\frac{k - (1-\alpha)N}{\beta N - k}} - \epsilon \right)^2 \cdot \frac{\mu_1^k e^{-\mu_1}}{k!}.
    \label{eq:MLE_MSE_exact}    
\end{equation}
(Here, we neglected the $N_1 > \beta N$ part, as it is extremely unlikely to occur for small separations when $\beta\approx1$.)

\begin{figure}
    \centering
    \includegraphics[width=1\linewidth]{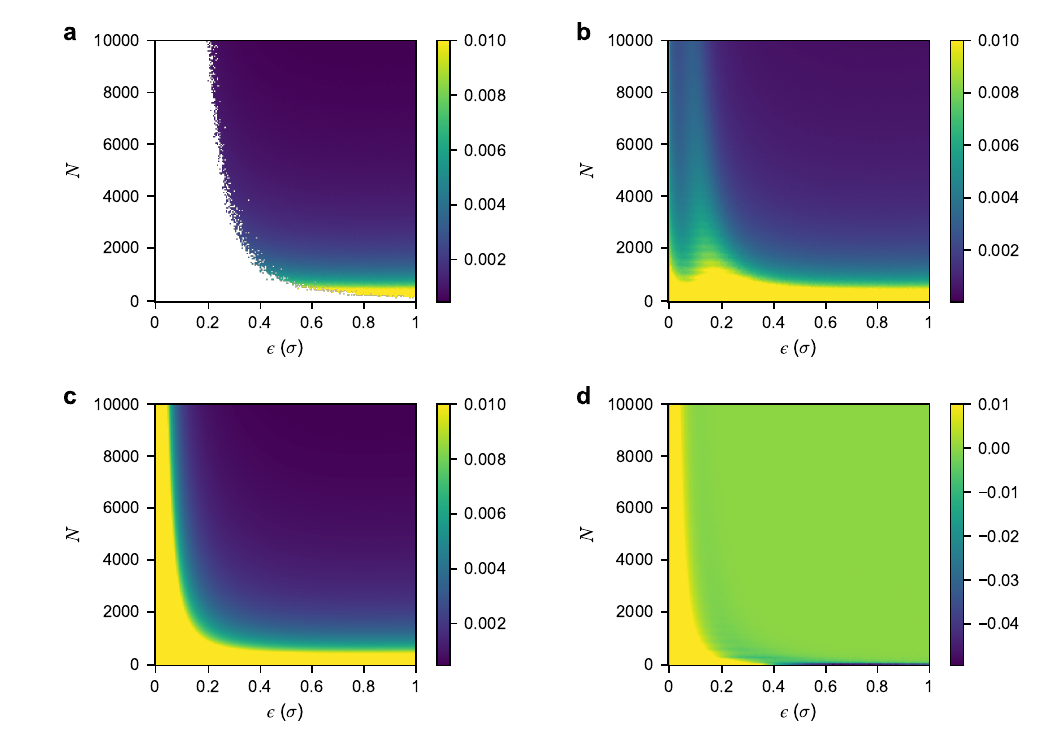}
    \caption{\textbf{Simulation results of the mean square error (MSE) and the unbiased Cramér-Rao lower bound (CRLB) for maximum likelihood estimation.} (a) Simulated MSE, where the white region indicates the onset of truncation due to the non-negativity constraint. (b) MSE with the non-negativity constraint applied. (c) The unbiased CRLB. (d) The difference between the CRLB and the MSE (CRLB $-$ MSE).}
    \label{fig:2D_MSE}
\end{figure}

The MSE simulated using our experimentally measured crosstalk is shown in Fig.~\ref{fig:2D_MSE} as a function of the separation $\epsilon$ and the number of photons $N$. In panel (a), the white region indicates where the MLE estimators begin to truncate negative estimates, and panel (b) displays the corresponding MSE with this truncation applied. Panel (c) presents the unbiased Cramér-Rao lower bound (CRLB). By taking the difference between the unbiased CRLB and the MSE (CRLB $-$ MSE), we obtain panel (d), which highlights a transition: for large separations, the MSE exceeds the unbiased CRLB (blue regions), but as the separation decreases and with larger bias, it drops below the CRLB (yellow regions).

\subsection{Quantum memory interaction}

To characterise the mappings between the signal mode and the spin-wave mode, we can describe the mapping using Green's functions \cite{Nunn_2008}, as the Raman memory equations are linear:
\begin{equation}
    B_\mathrm{stor}(z) =\int_{-\infty}^{+\infty}\mathrm{d}\tau\,K_1(z,\tau)S_\mathrm{in}(\tau), \label{eqn:Green_storage}
    \end{equation}
\begin{equation}
    S_\mathrm{out}(\tau) =\int_{0}^{L}\mathrm{d}z\,K_2(z,\tau)B_\mathrm{stor}(z),
\end{equation}
where $S_\text{in}(\tau)$ is the input signal mode, $B_\text{stor}(z)$ is the stored spin-wave mode, $S_\text{out}(\tau)$ is the retrieved optical mode. $K_{1}$ and $K_{2}$ are the corresponding Green's function components that characterize the storage and retrieval processes and contain information about the corresponding control fields, such as temporal modes and intensities.

To derive the memory efficiencies, we first need to define the number of excitations $N$ in each process.
\begin{equation}
    N_\text{in(out)}= \int_{-\infty}^{\infty}d\tau\left<S_\text{in(out)}^\dag(\tau)S_\text{in(out)}(\tau)\right>
    \label{eqn:Ni}
\end{equation}
\begin{equation}
    N_\text{stor}= \int_{0}^{L}dz\left<B_\text{stor}^\dag(z)B_\text{stor}(z)\right>.
    \label{eqn:excittaions2}
\end{equation}
where $N_\text{in(out)}$ is the average number of input (output) excitations in the optical field, and $N_\text{stor}$ is the average number of the stored spin-wave excitations in the medium. Thus, the storage, retrieval and total memory efficiencies are given as
\begin{equation}
    \begin{aligned}
        \eta_\mathrm{storage}   & =  N_\text{stor}/N_\text{in} = 1-N_\text{tran}/N_\text{in}\\
        \eta_\mathrm{retrieval} & =  N_\text{out}/N_\text{stor} = N_\text{out}/(N_\text{in}-N_\text{tran})\\
        \eta_\mathrm{total}     & =  N_\text{out}/N_\text{in}. 
    \end{aligned}
    \label{eq:efficiencies}
\end{equation}
where $N_\text{tran} = N_\text{in} - N_\text{stor}$ is the transmitted number of excitations. Experimentally, we used the transmitted photon counts to calculate the storage efficiency, as the number of excitations is not accessible.

The modal properties of the Raman memory can be understood by performing singular value decomposition on the Green functions $K_1$ and $K_2$ \cite{Nunn_2008}. For the storage process, $K_1(z,\tau)$ can be decomposed into 
\begin{equation}
    K_1(z,\tau) = \sum _ k \lambda_k \psi_k(z)  \phi_k^*(\tau)
\end{equation}
where $\lambda_k$ are the eigenvalues of the decomposition, sets of $\psi_k(z)$ and ${\phi_k(\tau)}$ represent the orthonormal bases of spin-waves and input temporal modes, respectively. In a single-mode quantum memory, only a specific input temporal mode $\phi_1(\tau)$ is efficiently coupled and stored as a corresponding spin-wave mode $\psi_1(z)$, while all orthogonal modes $\phi_k(\tau)$ for $k \neq 1$ pass through the medium unaffected. This selective interaction effectively allows the memory to function as a mode filter, storing one targeted signal while leaving others undisturbed. Crucially, the mode that is stored can be tuned by shaping the control pulse used in the Raman interaction. Adjusting the pulse’s shape, timing, and intensity modifies the Green function $K_1(z,\tau)$, which in turn determines the set of temporal modes ${\phi_k(\tau)}$ and their associated coupling strengths, represented by eigenvalues $\lambda_k$.

\subsection{Estimation results with different detected photon budgets}

\begin{figure}
    \centering
    \includegraphics[width=1\linewidth]{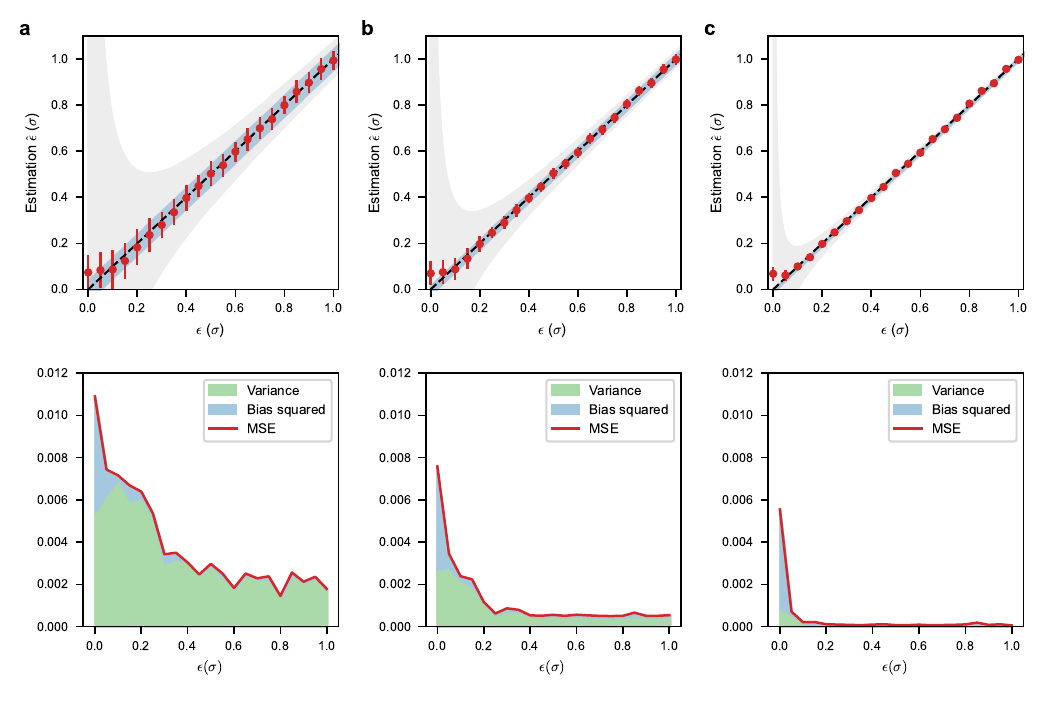}
    \caption{\textbf{Maximum likelihood estimation results for different total detection counts.} (a), (b), and (c) correspond to total detection counts of $N=$ $2\times10^3$, $10\times10^3$ and $100\times10^3$, respectively. The upper panels show the MLE estimators, along with the corresponding quantum bounds (blue region) and direct intensity measurement bounds (grey region). Error bars represent the estimator variances. The lower panels display the MSE (red line),  with contributions from bias squared (blue region) and variance (green region).}
    \label{fig:extra_results}
\end{figure}

To compare the performance of MLE estimation under different photon budgets, we present results for total detection counts of $N=$ $2\times10^3$, $10\times10^3$ and $100\times10^3$ in Fig.~\ref{fig:extra_results}. In all three cases, the MLE estimators closely follow the ground truth (dashed black line). The error bars fall below the intensity detection bound (grey region) but remain above the quantum limit (blue region), demonstrating improved precision over direct intensity (DI) measurements. As expected, the estimator variance decreases with increasing photon counts. The lower panels show the MSE (red line) along with its constituent components, as defined in Eq.~\ref{eq:MSE_bound} in the main text: variance (green area) and bias squared (blue area). The sum of these two contributions aligns well with the overall MSE of the estimators.

\subsection{Different retrieval times and retrieval modes}

\begin{figure}
    \centering
    \includegraphics[width=1\linewidth]{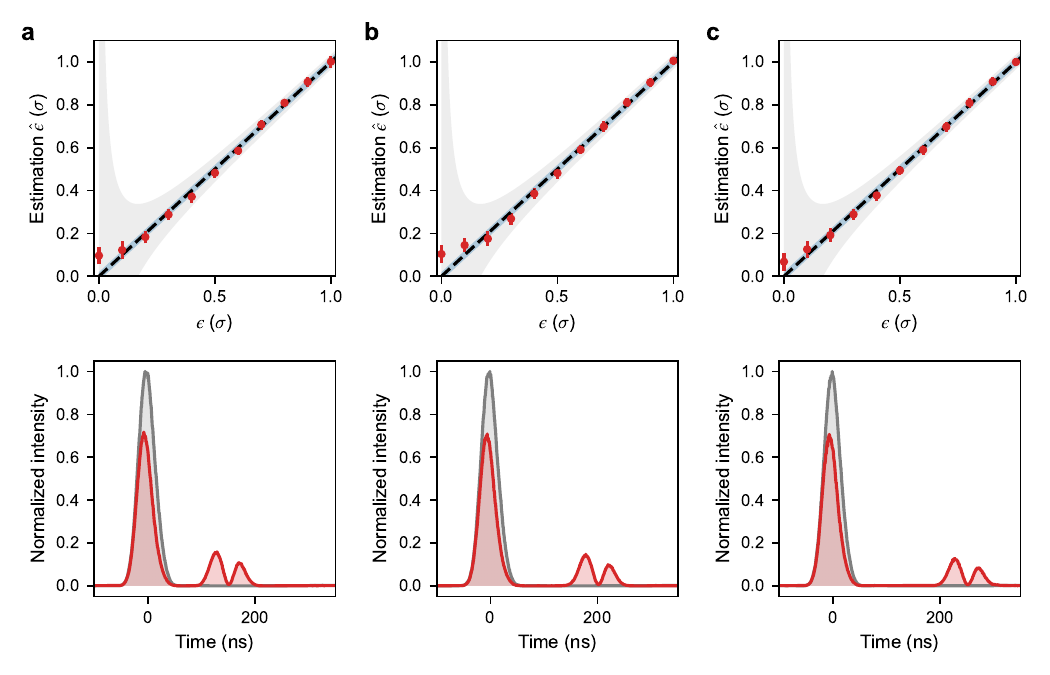}
    \caption{\textbf{Performance of MLE estimation at different retrieval times.} (a), (b), and (c) correspond to retrieval times of 150~ns, 200~ns, and 250~ns, respectively. In each case, the upper panels show the MLE estimates (red dots), along with the quantum bounds (blue region) and the direct intensity measurement bounds (grey region). The lower panels display the detected signal pulse sequences for two spectral lines with zero separation and zero relative phase. The input signal is shown in grey; the first red pulse indicates the leaked signal, and the second red pulse corresponds to the retrieved signal. The read-in control pulse is in the HG$_0$ mode, and the read-out control pulse is in the HG$_1$ mode.}
\label{fig:diff_time}
\end{figure}

Photonic quantum memories are designed to store photons over user-defined delay times. Raman quantum memories, in particular, offer additional functionality such as temporal mode conversion, where the stored signal can be retrieved into a temporal mode defined by the read-out control pulse. These combined capabilities of storage and mode conversion make Raman memories more versatile than previously demonstrated mode-filtering superresolution schemes, and could enable applications like distributed sensing networks, as discussed in the main text. Here, we demonstrate the superresolution performance of a Raman memory using different retrieval times and output temporal modes. 

The maximum storage time is generally limited by thermal diffusion in the warm vapour cell: atoms carrying the spin-wave coherence drift out of the interaction region (defined by the control beam size), rendering the stored signal unrecoverable. The typical coherence time is on the order of microseconds~\cite{Klein_2009}. In our experiments, we characterise the superresolution performance of the Raman memory at storage times of 150~ns, 200~ns, and 250~ns, as shown in Fig.~\ref{fig:diff_time}. These durations are chosen based on constraints from our delay fibre length and pulse sequence design. For all three retrieval times, the superresolution performance remains comparable, as illustrated in the upper panels of the figure. The lower panels show the raw detection counts from a representative experimental run, where the input signal had zero frequency separation and zero relative phase. The retrieved signals appear as the second red pulse, occurring precisely at 150~ns, 200~ns, and 250~ns, corresponding to retrieval using an HG\textsubscript{1} mode as the control read-out pulse.

\begin{figure}
    \centering
    \includegraphics[width=0.66\linewidth]{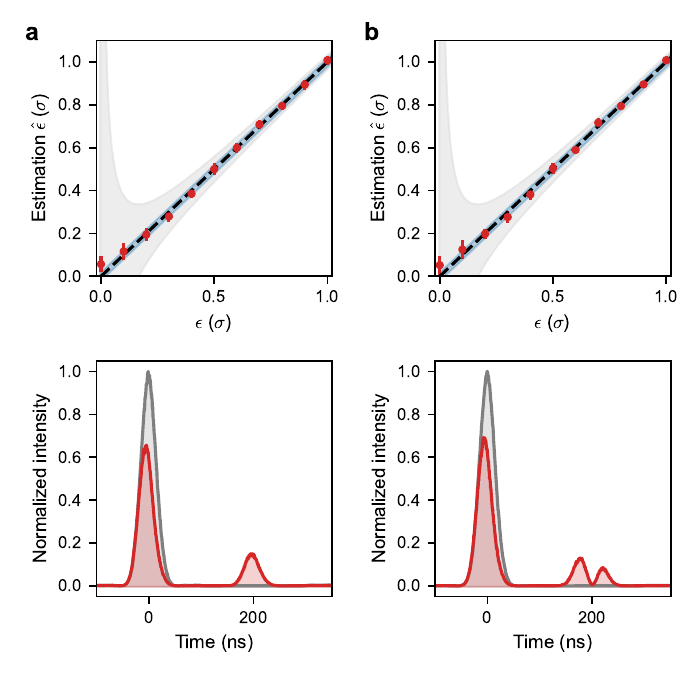}
    \caption{\textbf{Performance of MLE estimation with different retrieval modes.} (a) and (b) correspond to retrieval control modes of HG$_0$ and HG$_1$, respectively. The upper panels show the MLE estimates (red dots), along with the quantum bounds (blue region) and the direct intensity measurement bounds (grey region). The lower panels display the detected signal pulse sequences for two spectral lines with zero separation and zero relative phase. The input signal is shown in grey; the first red pulse indicates the leaked signal, and the second red pulse corresponds to the retrieved signal. The read-in control pulse is in the HG$_0$ mode.}
    \label{fig:diff_mode}
\end{figure}

We also performed the experiment using different retrieval modes to demonstrate the flexibility in choosing the read-out control pulse. The results are shown in fig.~\ref{fig:diff_mode}, where the control read-out pulses are set to HG\textsubscript{0} and HG\textsubscript{1}, respectively. The superresolution performance remains comparable in both cases, and the retrieved signals (the second red pulse) clearly exhibit the corresponding temporal mode shapes.

\subsection{Comparison of time–frequency super-resolution sensor architectures}

Table~\ref{tab1} presents a detailed comparison of frequency-domain super-resolution schemes~\cite{Donohue_2018, Mazelanik_2022,Lipka_2024}, evaluating their underlying principles, programmability, complexity, bandwidth, and precision enhancement. While PuDTAI (GEM) and Raman protocols both utilise quantum memories, the GEM architecture relies on mapping frequency components to longitudinal spatial positions via magnetic field gradients. This mechanism inherently restricts the system to the ultra-narrowband regime, while necessitating cumbersome magnetic field gradients and laser-cooling infrastructure. Pushing to a larger bandwidth requires a steeper gradient to span a wider detuning range across a fixed-length medium. Because the total optical depth is finite, this effectively reduces the optical depth per unit frequency, causing storage efficiency to drop as bandwidth increases. At the same time, the steeper gradient compresses the spatial mapping, shrinking the spatial separation between frequency components and making the stored coherence more susceptible to blurring from atomic motion. These physical constraints are compounded by engineering limits, since steeper gradients demand higher currents and more coil turns, increasing inductance and complicating the rapid gradient reversal required for retrieval. Furthermore, PuDTAI and SUSI are restricted to a fixed basis of symmetric and antisymmetric modes. Conversely, while QPG offers high programmability, it lacks storage capabilities and operates in a very high bandwidth regime ($> 100$~GHz).

\begin{table}
\caption{Comparison of time–frequency super-resolution sensor architectures}
\label{tab1}
\begin{tabularx}{\textwidth}{@{}l>{\raggedright\arraybackslash}X
                               >{\raggedright\arraybackslash}X
                               >{\raggedright\arraybackslash}X
                               >{\raggedright\arraybackslash}X@{}}
\toprule
\textbf{Scheme} & \textbf{QPG} & \textbf{PuDTAI} & \textbf{SUSI} & \textbf{Raman} \\
\midrule
\textbf{Principle} & Sum-frequency generation with shaped pulses & Time-inversion interferometry with spatial-spectral mapping & Spectral-domain inversion interferometry & Stimulated Raman scattering with pulse shaping \\
\addlinespace
\textbf{Programmability} & Arbitrary mode & Fixed (symmetric/ antisymmetric) & Fixed (symmetric/ antisymmetric) & Arbitrary mode \\
\addlinespace
\textbf{Storage \& Conversion} & No & Storage & No & Yes \\
\addlinespace
\textbf{Complexity} & Low (nonlinear waveguide, spatial-light modulator pulse shaping) & High (cold atoms, magnetic gradients, AC-Stark modulation) & Medium (interferometer stabilisation, electro-optic lenses) & Low (warm vapour, electro-optical modulator pulse shaping) \\
\addlinespace
\textbf{Bandwidth} & 100s of GHz to THz & kHz to low MHz & 10s of GHz to 100s of GHz & 100s of kHz to low GHz \\
\addlinespace
\textbf{Precision enhancement} & $\sim 24$ & $\sim 20$ & $\sim 2.13$ & $\sim37$ \\
\bottomrule
\end{tabularx}
\end{table}

\subsection{Sensing applications in Doppler LiDAR}

The mode-selective capability of our Raman quantum memory platform offers a distinct advantage for remote sensing applications, particularly in Doppler Light Detection and Ranging (LiDAR). In this section, we outline the operational principles of Doppler LiDAR and discuss how our platform enables high-precision velocity and range estimation.

Doppler LiDAR systems generally fall into two categories: coherent~\cite{Frehlich_1997} and incoherent~\cite{McGill_1997} detection. Coherent Doppler LiDAR relies on heterodyne detection, where the backscattered signal is mixed with a local oscillator (LO) to extract the Doppler frequency shift. While offering shot-noise-limited sensitivity, coherent Doppler LiDAR requires a narrow-linewidth laser, spatial mode matching between the signal and a phase-locked LO, and is susceptible to performance degradation from atmospheric turbulence and speckle noise. On the other hand, incoherent Doppler LiDAR typically utilises passive optical filters (such as Fabry-Pérot etalons or molecular absorption cells) with steep transmission slopes. The Doppler frequency shift is converted into an intensity variation as the signal moves along the filter's transmission edge. However, its performance and precision are fundamentally limited by the steepness of the optical filters and the trade-off between dynamic range and sensitivity.

Our Raman memory-based platform provides a high-precision, mode-selective measurement in an incoherent way that overcomes the above limitations. By performing mode-selective storage, our method offers a distinct advantage over static edge filters: the spectral properties of the filter are directly defined by the optical control field, allowing the measurement basis (centre frequency and bandwidth) to be flexible and dynamically programmable. This mode-selectivity also enhances noise resilience. Unlike conventional incoherent LiDARs that integrate background noise throughout the filter window, our system functions as a temporal filter, effectively rejecting any noise components that are temporally or spectrally orthogonal to the control mode. Moreover, unlike coherent LiDAR, our approach does not require a phase-stable local oscillator, making it resilient to noise from turbulence and phase decorrelation. 

By leveraging single-photon detection and optimal mode filtering, our platform achieves fine resolution in low signal-to-noise ratio (SNR) regimes, where conventional methods fail. A compelling application is remote vibrometry, applicable to scenarios such as laser ultrasonic testing for defect detection in industrial composites and non-contact photoacoustic sensing in biomedical imaging, where a stationary target exhibits surface oscillations in the MHz regime. For instance, a target vibrating with multiple acoustic frequencies will generate a return signal composed of closely spaced spectral lines, enabling high-sensitivity at MHz-GHz high-speed acoustic monitoring even in the presence of turbulence. In the following sections, we detail specific scenarios where this technology can be applied as a Doppler LiDAR.

\subsubsection{Single target velocity measurement}

Consider a scenario where a pulsed laser with spectral width $\sigma$ is emitted and reflected by a single moving target, acquiring a Doppler frequency shift $\delta$ proportional to its velocity. To measure this frequency shift, we employ a mode-selective measurement that directly infers the frequency shift from the detected photon counts. Through a pre-calibration of the storage efficiency $\eta$ as a function of signal central frequency, the unknown shift $\delta$ is retrieved by comparing the measured photon statistics of the filtered signal. Crucially, this platform fills a technological gap in the MHz to GHz bandwidth regime, a domain where conventional direct intensity techniques (such as grating or Fourier transform spectrometers) are inaccessible. Furthermore, unlike passive filters, the spectral position and bandwidth of our acceptance mode can be dynamically adjusted, effectively allowing us to select the optimal ``edge" to maximise sensitivity for the expected velocity range.

\subsubsection{Joint estimation of range and velocity}
Conventional Doppler LiDAR systems relying on very narrowband lasers often provide poor resolution in range estimation due to the fundamental Fourier limit. By operating with high-bandwidth pulses, our platform is capable of simultaneously estimating range and velocity with high resolution.  In this scenario, the return signal is characterised by both a time delay $\tau$ (encoding range) and a frequency shift $\delta$ (encoding velocity). Since the Raman interaction is strictly mediated by the control field, the storage process effectively projects the input signal onto a specific time-frequency mode basis defined by the control pulse's temporal envelope and central frequency. This allows for the optimisation of the measurement to maximise Fisher Information for both parameters.

\subsubsection{Resolving two or more objects}

A more complicated scenario involves two or more targets moving at slightly different velocities~\cite{Huang_2021}. The return signal becomes an incoherent mixture of two or more time-frequency modes: $\rho = \sum_i p_i |\psi(t-\tau_i, \omega-\delta_i)\rangle \langle \psi(t-\tau_i, \omega-\delta_i)|$. This scenario maps directly to resolving closely spaced spectral lines addressed in our work. When the velocity difference is sub-Rayleigh and the two spectral lines overlap, conventional systems struggle to resolve the two velocities. By shaping the control field, we can perform an optimised mode-selective measurement for resolving the range and velocity of two or more objects. In many practical LiDAR and vibrometry scenarios, decoherence occurs when reflecting off rough surfaces (diffuse scattering) or propagating through atmospheric turbulence, where the relative phases between reflections from different scattering centres are randomised. Consequently, the interference terms average to zero, and the return signal is accurately described by the incoherent mixed state $\rho$. Our method is therefore particularly advantageous for sensing diffuse targets, where traditional coherent detection (heterodyne) is often degraded by speckle noise and wavefront distortions.

\subsection{Sources of mode crosstalk}

In this work, we employed an imperfect two-mode projection method to generate our estimators, incorporating all imperfections into the parameters $\alpha$ and $\beta$ in the perturbation matrix $\boldsymbol{M}$ (Eq.~\ref{eq:crosstalk_matrix} in the main text). Below, we discuss the primary sources of crosstalk that contribute to the mode crosstalk and the bias in the raw estimator.

The primary source of temporal mode crosstalk arises from the memory’s storage interaction. Although the memory operates approximately in a single-mode regime, increasing the control field strength induces stronger AC Stark shifts. This leads to the coupling of orthogonal signal modes into the spin wave, distorting the ideal projection probabilities and introducing bias. Fig.~\ref{fig:sim_diff_eff} illustrates how different storage efficiencies (i.e. different control read-in pulse power) affect the superresolution performance and estimation precision relative to DI methods.

Panel (a) shows the raw estimators $\hat{\epsilon} = 4\sqrt{N_1/N_0}$ for different storage efficiencies. As storage efficiency increases, so does crosstalk, resulting in greater bias in the raw estimators. In panel (b), we apply maximum likelihood estimation and plot the corresponding unbiased CRLB for the two-mode demultiplexing method, alongside the quantum CRLB and the CRLB for DI methods. Higher storage efficiency leads to an increase in the CRLB, resulting in reduced estimation precision. This trend is further quantified in panel (c), where the precision improvement factor, $\mathcal{F}/\mathcal{F}_\text{DI}$, is plotted as a function of separation $\epsilon$. As the storage efficiency decreases, the superresolution parameter increases nearly logarithmically. However, very high precision improvement using very low storage efficiencies is impractical in experiments, as fewer signal photons are retrieved, constant background and control field leakage reduce the signal-to-noise ratio, potentially increasing the observed crosstalk and estimation bias.

Another major limiting factor contributing to excess crosstalk in our experiment is control field leakage, which introduces a constant background noise. Due to the small frequency separation (9.2 GHz) between the signal and control fields, completely filtering out the control field is technically challenging. As a result, residual control photons are detected and contribute equally to both HG$_0$ and HG$_1$ projection measurement counts, adding uniform noise across all separations that distorts the projection probabilities.

To quantify this effect, we simulate how constant background counts influence the bias of the raw estimators and the precision improvement of MLE estimators. In this simulation, we fix the storage efficiency at 40\% and introduce a constant leakage level ranging from 0 to 2\% of the total detected counts. The results are shown in fig.~\ref{fig:sim_leak}. Panel (a) shows that increasing control leakage leads to a larger bias in the raw estimator, similar to the effect of increasing storage efficiency. The rate of bias increase slows as the leakage grows uniformly. Panel (b) presents the corresponding precision improvement, which diminishes with higher background noise due to the increasing bias.
\begin{figure}
    \centering
    \includegraphics[width=1\linewidth]{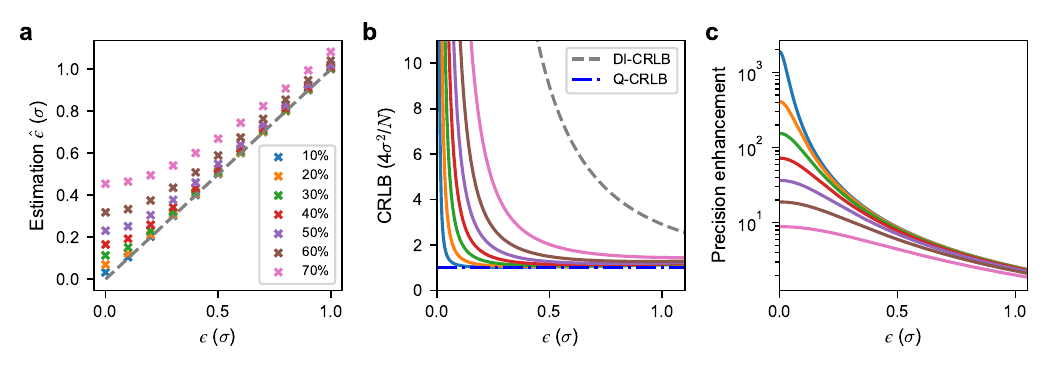}
    \caption{\textbf{Effect of storage efficiency on superresolution performance.} (a) Raw estimators $\hat{\epsilon} = 4\sqrt{N_1/N_0}$ for various storage efficiencies, showing increased bias at higher efficiencies due to stronger temporal mode crosstalk. (b) Unbiased CRLB for the two-mode demultiplexing method at different storage efficiencies, compared to the quantum CRLB and the CRLB for the direct intensity measurement method. (c) Precision enhancement $\mathcal{F}/\mathcal{F}_{\text{DI}}$ as a function of separation $\epsilon$, demonstrating logarithmic improvement for small separations.}
    \label{fig:sim_diff_eff}
\end{figure}

\begin{figure}
    \centering
    \includegraphics[width=0.66\linewidth]{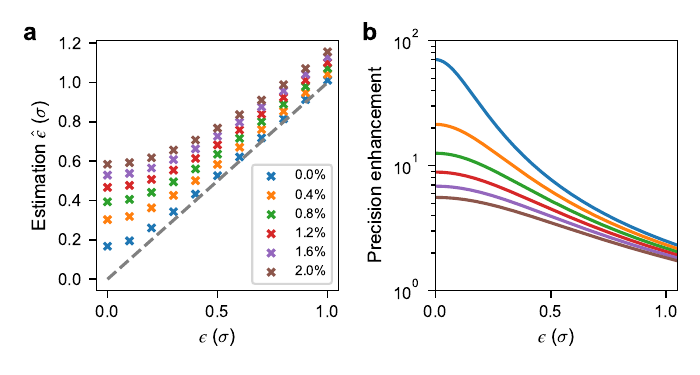}
    \caption{\textbf{Effect of control leak counts on estimation bias and precision.} (a) The bias in the raw estimators increases with control leakage, expressed as a percentage of the total detection counts. (b) Precision enhancement of the mode filtering method decreases as background noise increases. Storage efficiency is fixed at 40\%.}
    \label{fig:sim_leak}
\end{figure}

Additional sources of mode crosstalk include detector dark counts, signal background from incomplete extinction of the electro-optic modulator, and four-wave mixing (FWM) noise generated during the Raman interaction. In our experiment, we minimised the signal background using a Pockels cell and characterised both the dark counts and residual background, which were subtracted from the measured data. Furthermore, we applied appropriate detunings to the Raman transition to naturally suppress FWM noise, as described in Ref.~\cite{Thomas_2019}. Thus, the contribution of these sources is minimal.

\subsection{Precise pulse carving}

\begin{figure}
    \centering
    \includegraphics[width=0.66\linewidth]{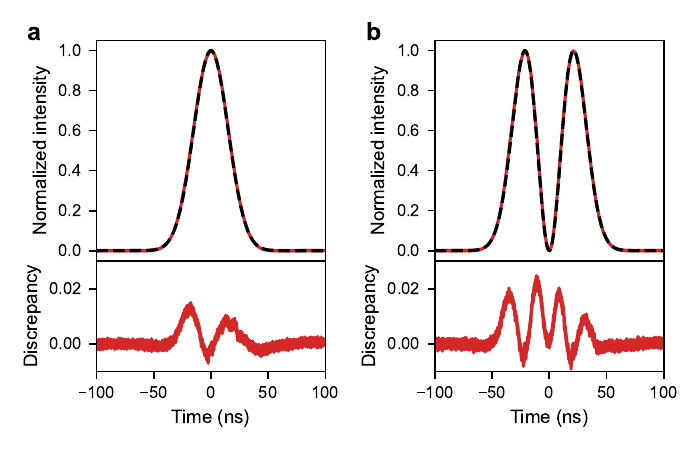}
    \caption{\textbf{Intensity profiles of experimental HG control pulses.} (a) and (b) show the measured intensity profiles of the HG$_0$ and HG$_1$ control pulses, respectively. In the upper panels, the experimentally measured normalised profiles (solid red) are compared with the theoretical normalised HG profiles (dashed black). The lower panels display the discrepancies between the experimental and theoretical profiles.}
    \label{fig:exp_HG_pulses}
\end{figure}

To improve our pulse carving system, we applied frequency response correction to our electronic system. In a linear time-invariant (LTI) system, where our electronic system can be approximated as such, the electric signal output $h(t)$ of the system to an arbitrary electric signal input function $x(t)$ can be approximately described by a linear response function $R(t-t')$ as
\begin{equation}
h(t) \approx \int_{-\infty}^{t}dt'R(t-t')x(t').
\label{eq:response}
\end{equation}
In the frequency domain, the frequency response function $\tilde{R}(\omega)$ can be calculated as,
\begin{equation}
\tilde{R}(\omega) = \frac{\tilde{h}(\omega)}{\tilde{x}(\omega)}.
\label{eq:responsefunction}
\end{equation}
Once the response function is known, a target signal $h'(t)$ can be precisely generated by using the response-corrected input signal $x'(t)$:
\begin{equation}
x'(t) = \mathcal{F}^{-1}\left[\frac{\mathcal{F}(h'(t))}{\tilde{R}(\omega)}\right].
\label{eq:correction}
\end{equation}
where $\mathcal{F}$ denotes the Fourier transform and $\mathcal{F}^{-1}$ denotes the inverse Fourier transform.

To demonstrate our high-fidelity optical pulse carving system, we present the averaged and normalised intensity profiles of the HG$_{0}$ and HG$_{1}$ optical control pulses, as measured by a photodiode and an oscilloscope (1024 averages). fig.~\ref{fig:exp_HG_pulses}(a) and (b) show the experimentally measured pulse shapes (red) compared against the theoretical intensity distributions (black dashed lines). The discrepancies between the theoretical and experimental data are plotted in the lower panels. The deviation between the measured and theoretical profiles is small, typically not exceeding 2\% relative to the maximum.

\end{widetext}
\end{document}